# ORDERMOE: AN EXPERT SIMILARITY DRIVEN DISTRIBUTED EDGE MOE INFERENCE

Xin Yuan, Ning Li, Quan Chen, Wenchao Xu, Athanasios V. Vasilakos, Song Guo, *Fellow, IEEE*, Haijun Zhang, *Fellow, IEEE*

**Abstract—Although mixture-of-experts, MoE, models have been increasingly adopted to scale large language models with moderate computation cost, it remains challenging to deploy MoE inference over resource-constrained and bandwidth-limited edge infrastructures. Existing distributed MoE serving methods mainly rely on exact expert placement, caching, replication, or communication scheduling, while overlooking the functional similarity among experts, which provides an opportunity to reduce cross-server token transmission. Therefore, this paper introduces a similarity-aware expert allocation and distributed deployment framework, dubbed OrderMoE, which aims to accelerate edge MoE inference while balancing inference latency, communication overhead, server workload, and inference quality. OrderMoE first constructs an expert similarity model based on router-induced logits representations and partitions experts in each MoE layer into multiple similarity groups. Then, it develops a similarity-aware expert grouping and deployment strategy to improve local similarity coverage across edge servers. Since reducing remote expert invocation and preserving exact inference quality are conflicting objectives, OrderMoE further designs a quality-aware and trajectory-aware runtime server-expert selection algorithm to decide whether a token should invoke its remote target expert or use a feasible local substitute expert. Experimental results on a real distributed edge testbed show that OrderMoE significantly reduces average latency, tail latency, cross-server traffic, and remote expert invocation ratio, while introducing only small and controllable inference quality degradation.**

## I. INTRODUCTION

Large language models (LLMs) have demonstrated remarkable capabilities in a wide range of intelligent services, such as mobile assistants, augmented reality, autonomous driving, real-time content generation, etc. [1]. However, the huge parameter size and computation complexity of LLMs make it difficult to deploy them directly on resource-constrained end devices and edge servers [2]. To address this issue, edge intelligence has become a promising paradigm, in which model inference is collaboratively executed by end devices and edge infrastructure to reduce latency, bandwidth consumption, and privacy risks. Among various LLM architectures, mixture-of-experts (MoE) models are particularly attractive for edge intelligence because they activate only a sparse subset of experts for each token and thus provide a favorable tradeoff between model capacity and computation cost [3].

For deploying LLMs at the edge, existing studies mainly have two technical directions. The first direction is model compression, such as quantization, pruning [4], distillation [5], low-rank approximation [6], which reduces the model size and computation cost before deployment. However, model compression may cause inference accuracy degradation, and for very large LLMs, the compressed model may still exceed the memory and computation capacity of a single edge server [7]. The second direction is distributed deployment, in which a large model is partitioned and collaboratively executed across multiple servers [8]. Compared with compression-only approaches, distributed deployment is more suitable for supporting large-scale LLMs at the edge, because the memory and computation burdens can be shared by multiple edge servers [9]. Therefore, collaborative deployment among edge servers becomes necessary when a single edge server cannot accommodate the complete LLM or provide sufficient inference capability.

The distributed deployment of large models has been widely investigated in cloud data centers, where model partitions can be executed across multiple servers to improve serving efficiency [10]. In cloud environments, servers are usually connected by dedicated high-speed interconnects, such as RDMA [11], InfiniBand [12], etc., and thus the communication overhead caused by cross-server execution can often be effectively controlled. However, directly applying the approaches of cloud distributed deployment to edge intelligence is not suitable. On the one hand, different from cloud data centers, geographically distributed edge servers are commonly

This paragraph of the first footnote will contain the date on which you submitted your paper for review, which is populated by IEEE. This work was supported in part by the grant from NSFC Grant no. 62571156, 62101159, 52475009, NSF of Shandong Grant no. ZR2021MF055, ZR2025QC666, the Research Grants Council of Hong Kong under the Areas of Excellence scheme grant AoE/E-601/22-R, and also the Opening Project of the Key Laboratory of Advanced Manufacturing and Intelligent Technology (Ministry of Education) at Harbin University of Science and Technology (KFKT202306).
*(Corresponding author: Ning Li).*

Xin Yuan is with the School of Ocean Engineering, Harbin Institute of Technology, Heilongjiang, China (e-mail: xin.yuan@hit.edu.cn).

Ning Li is with the School of Computer Science and Technology, Harbin Institute of Technology, Heilongjiang, China (e-mail: li.ning@upm.es).

Quan Chen is with the School of Computer Science and Technology, Guangdong University of Technology, Guangdong, China (e-mail: quanchen89@126.com)

Wenchao Xu is with the Division of Integrative Systems and Design, Hong Kong University of Science and Technology, Hong Kong (e-mail: wenchaoxu@ust.hk)

Athanasios V. Vasilakos is with the Center for AI Research (CAIR), University of Agder (UiA), Grimstad, Norway (e-mail: th.vasilakos@gmail.com).

Song Guo is with the Department of Computer Science and Engineering, Hong Kong University of Science and Technology, Hong Kong (e-mail: songguo@cse.ust.hk)

connected through the ordinary Internet rather than specialized high-bandwidth interconnects [13]. Thus, the available bandwidth among edge servers is usually much smaller and the transmission latency is higher; moreover, the dynamic network condition makes this issue more serious. On the other hand, the memory capacity, computational power, and real-time workload of edge servers vary significantly, which further complicates the distributed deployment of LLMs at the edge [14].

These edge-specific characteristics make distributed MoE deployment particularly challenging. In most existing related works, the activated expert in MoE is treated as a fixed computation unit. Once the routed expert of a token is not deployed on the local server, the token representation has to be transmitted to a remote server for exact expert execution. Since MoE routing decisions are input-dependent, such exact expert access may trigger frequent cross-server transmission during token inference. This design may be acceptable in cloud data centers with high-speed interconnects, but it can introduce considerable delay and unstable performance in decentralized edge environments. More importantly, most existing approaches focus on improving the efficiency of accessing the exact remote expert, but do not investigate whether the remote invocation itself can be avoided. Therefore, reducing cross-server token transmission and keeping token computation local still remain challenging and attractive problems in distributed edge MoE inference.

Fortunately, recent observations indicate that experts within the same MoE layer are not completely isolated. Instead, many same-layer experts exhibit functional similarity, and such similarity can be measured by the row vectors of the router logits matrix, which reflects the routing behavior of experts over input embeddings [15]. Moreover, the similarity among experts is also related to the depth of MoE layers. In general, experts in deeper layers tend to be more similar to each other, while experts in shallower layers usually have lower similarity and stronger specialization [16]. This phenomenon provides important guidance for similarity-aware expert grouping and deployment. Based on this observation, if a target expert required by a token is absent from the local edge server, it may be possible to use a highly similar local expert to perform substitute computation, instead of transmitting the token to a remote edge server for exact expert execution. In this manner, the distributed deployment problem of MoE inference can be transformed from exact remote expert access to similarity-aware local substitute execution. To the best of our knowledge, this is the first work that exploits same-layer expert similarity for distributed collaborative deployment across multiple edge servers, with the purpose of reducing cross-server token transmission and keeping token inference local as much as possible.

Nevertheless, exploiting similar experts for distributed edge deployment is not easy. First, the experts should be organized according to their same-layer similarity, and the layer-dependent similarity characteristics should be considered during expert grouping. For example, deeper layers may allow larger similarity groups due to higher expert similarity, while shallower layers may require more conservative grouping because their experts are less similar. Second, under the limited memory and computation resources of edge servers, determining which experts should be deployed on which server is difficult, because the deployment strategy should maximize local substitutability while minimizing expected cross-server communication. Third, local substitution of a remote expert may reduce communication delay, but it may also introduce inference quality loss. Therefore, an effective distributed deployment framework should carefully balance communication reduction and inference quality preservation. Finally, in practical edge environments, the servers are heterogeneous in memory capacity and computation capability, which further complicates the design of similarity-aware expert deployment.

To address these issues, in this paper, a similarity-aware expert grouping and distributed deployment framework for edge MoE inference, abbreviated as OrderMoE, is proposed. Specifically, for each MoE layer, OrderMoE first constructs an expert similarity model based on router-induced logits representations; then it partitions the experts into multiple similarity groups. After that, a similarity-aware expert grouping and distributed deployment strategy is designed to distribute similar experts across multiple edge servers, such that each server maintains a local subset of representative experts from different similarity groups. Based on this deployment, when the routed target expert of a token is not available locally, the token can be processed by a highly similar local expert instead of being forwarded to a remote edge server. Furthermore, to control the tradeoff between communication reduction and inference quality, a quality-aware token trajectory and server-expert selection strategy is introduced to determine when local substitute execution should be adopted and when remote exact expert invocation is still necessary. In this way, the proposed framework reduces cross-server token transmission over bandwidth-limited and dynamic edge interconnections, and improves the efficiency of distributed edge MoE inference.

The main contributions of this paper are summarized as follows.

- We formulate the distributed edge MoE inference problem by considering expert deployment, expert similarity, substitute execution, and token trajectory. Different from existing methods, the proposed formulation allows a token to use quality feasible substitute experts and dynamically updates the token location after each MoE layer.
- We propose a similarity-aware expert grouping and deployment strategy. Experts in the same layer are grouped according to their functional similarity, and representative experts are deployed across edge servers to improve similarity group coverage. In addition, important experts are redundantly deployed according to their activation frequency and similarity role, so that the remote expert invocation probability can be reduced.
- We design a token trajectory aware online server-expert selection algorithm. The algorithm jointly considers transmission delay, computation delay, substitution quality loss, server workload, and future token movement cost. It decides not only whether a token should be locally substituted or remotely executed, but also whether the token should stay on the remote execution server for following layers.
- We implement OrderMoE on a real distributed edge testbed. Experimental results show that OrderMoE can significantly

reduce average latency, tail latency, cross-server communication traffic, and remote exact expert execution ratio, while introducing only small and controllable inference quality degradation.

The rest of this paper is organized as follows. Section II reviews the related work. Section III presents the system model and problem formulation. Section IV introduces the similarity-aware expert grouping and deployment strategy. Section V presents the token trajectory aware runtime inference algorithm. Section VI evaluates the performance of OrderMoE. Finally, Section VII concludes this paper.

## II. Related Works

In this section, we review the existing studies related to this paper from three aspects: 1) collaborative inference and distributed deployment of large models at the edge, 2) MoE serving optimization and communication reduction, and 3) expert similarity, clustering, and structural reorganization in MoE models. Although these studies have significantly improved the efficiency of large-model inference from different perspectives, they cannot address the problem discussed in this paper effectively, because they do not investigate similarity-aware distributed expert deployment across multiple edge servers to reduce cross-server token transmission.

### *A. Collaborative Inference and Distributed Deployment of Large Models at the Edge*

The deployment and inference acceleration of large models in edge environments have been investigated in many previous studies. Existing works mainly focus on model partitioning, collaborative inference, task offloading, and resource allocation to reduce the end-to-end latency and improve the efficiency of resource-constrained edge intelligence. For instance, Mohammed et al. [17] investigate adaptive DNN partitioning and offloading for distributed inference acceleration, while PArtNNer [18] proposes a platform-agnostic adaptive edge-cloud DNN partitioning method to minimize end-to-end latency. Recently, a comprehensive survey in [19] further summarizes the modeling, optimization objectives, and toolchains for cooperative DNN inference in edge intelligence, showing that collaborative inference has become a major paradigm for edge AI systems.

With the rapid development of LLMs, recent studies have extended collaborative inference from conventional DNNs to large generative models. Edge-LLM [20] proposes a collaborative framework for LLM serving in edge computing. EdgeShard [21] investigates collaborative edge inference for LLMs, and Communication-Efficient Distributed On-Device LLM Inference Over Wireless Networks [22] explores distributed inference across multiple devices by partitioning model tensors. Jupiter [23] and Birds in Cages [24] also study collaborative inference or inference allocation for LLMs on edge devices or edge servers. In addition, Cloud-Edge-End Collaborative Inference in Mobile Networks [25] discusses the challenges and solutions for collaborative inference of large models in mobile networking scenarios. More related works can be found in [26][27][28].

However, these works are fundamentally different from this paper. First, they mainly consider model partitioning, tensor partitioning, or inference assignment, and do not exploit the internal expert similarity of MoE models as a deployment primitive. Second, most existing collaborative inference designs are based on exact model execution, i.e., when a model component is assigned to a remote node, the corresponding data must still be transmitted to that node for exact execution. Third, although these studies improve the efficiency of distributed inference, they do not investigate how to reduce cross-server communication by replacing remote exact expert execution with local similar expert execution. Therefore, they cannot address the problem studied in this paper.

### *B. MoE Serving Optimization and Communication Reduction*

The efficiency of MoE inference has attracted increasing attention in recent years, because the sparsity of MoE models introduces not only computation efficiency but also new communication and memory challenges. In particular, when the activated experts of tokens are distributed across different devices or servers, expert dispatch and communication may become a dominant bottleneck.

Several existing studies optimize MoE inference through expert offloading, caching, prefetching, replication, and communication scheduling. AdapMoE [29] studies adaptive sensitivity-based expert gating and management for efficient MoE inference. SlimCaching [30] investigates edge caching of MoE models for distributed inference. Diff-MoE [31] proposes priority-driven differential expert caching to improve batched MoE inference efficiency. PROBE [32] further improves MoE inference by co-balancing computation and communication through predictive prefetching, dynamic expert replication, and token assignment. MegaScale-Infer [33] studies disaggregated expert parallelism for efficient large-scale MoE serving. These studies demonstrate that communication-aware expert management is important to MoE inference efficiency.

Some recent studies also investigate distributed or wireless MoE deployment at the edge. WDMoE [34] proposes a wireless distributed MoE architecture for LLMs. MoE2 [35] studies collaborative inference optimization for edge LLMs. In addition, EdgeShard [21] and related edge collaborative inference systems also partially touch the communication bottleneck of distributed sparse model serving.

However, the above studies still differ from this paper in several key aspects. First, their objective is mainly to improve the efficiency of accessing exact target experts by better caching, replication, prefetching, dispatch, or scheduling. In contrast, this paper investigates whether the remote target expert invocation itself can be avoided. Second, most existing studies assume that the routed target expert is the only valid execution target, and therefore optimize the communication path to that expert. In this paper, we exploit the similarity among same-layer experts and investigate how to keep token computation local whenever possible. Third, although some of the above studies consider distributed edge deployment, they do not design a similarity-aware deployment strategy that deliberately disperses representative similar experts across multiple edge servers to support local substitute execution. Therefore, they cannot solve the problem studied in this paper.

### *C. Expert Similarity, Clustering, and Structural Reorganization in MoE Models*

Another line of related work attempts to exploit structural redundancy or similarity in MoE models. Recent studies have shown that experts in MoE models may exhibit functional similarity, parameter redundancy, or clustered activation behavior. Based on this observation, some works reorganize experts to improve routing efficiency, reduce parameter redundancy, or support model compression.

Breaking the MoE LLM Trilemma [36] introduces dynamic expert clustering with structured compression to reduce load imbalance, parameter redundancy, and communication overhead. The authors regroup experts using parameter and activation similarity, and then perform hierarchical routing and shared structural decomposition within each cluster. In addition, recent surveys on MoE inference optimization [37] and large-model serving systems [38] summarize expert clustering, shared experts, communication bottlenecks, and structural optimization as important research directions. These studies indicate that expert similarity and clustering are valuable for improving the scalability and efficiency of MoE models.

Nevertheless, these studies are still different from this paper. First, they mainly use similarity or clustering for model restructuring, routing simplification, compression, or training-time/system-level optimization, rather than for distributed collaborative deployment over multiple edge servers. Second, their focus is not on the unique characteristics of edge environments, such as decentralized deployment, Internet-based inter-server communication, and strong heterogeneity of edge resources. Third, although clustering may reduce communication indirectly, these studies do not explicitly investigate the problem of local similar-expert substitution to reduce cross-server token transmission. Therefore, they cannot solve the problem considered in this paper.

### *D. Summary of Differences*

In summary, the existing studies on collaborative inference at the edge mainly focus on model partitioning, task offloading, and resource allocation [17-28]. The existing studies on MoE serving optimization mainly improve the efficiency of exact expert access through caching, prefetching, replication, and communication scheduling [29-35]. The studies on expert similarity or clustering mainly focus on structural reorganization, compression, or routing simplification [36-38]. Different from all of them, this paper investigates a new problem: how to exploit same-layer expert similarity to support distributed collaborative deployment across multiple edge servers, such that token computation can be kept local whenever possible and cross-server transmission over bandwidth-limited and dynamic edge interconnections can be reduced.

## III. Network Model and Problem Statement

In this section, the network model, distributed MoE deployment model, expert similarity and grouping model, token trajectory aware inference model, inference delay model, and substitution quality model that will be used in this paper are introduced. Based on these models, the problem that will be solved in this paper is described in detail.

### *A. Network Model*

As shown in Fig.1, we consider a distributed edge intelligence system consisting of multiple mobile users and edge servers. Let $\mathcal{D} = \{d_1, d_2, \dots, d_U\}$ denote the set of mobile users, where $U$ is the number of users. Each mobile user can generate inference requests for intelligent services. These inference requests are sent to nearby edge servers through wireless access networks. Let $\mathcal{S} = \{S_1, S_2, \dots, S_M\}$ denote the set of edge servers, where $M$ is the number of edge servers. For mobile user $d_i$, its associated edge server is denoted by $S_{a(i)}$, where $a(i) \in \{1,2,\dots,M\}$. The associated edge server receives the user request and serves as the initial execution server of the inference task. In this paper, we mainly focus on the distributed deployment and inference process among edge servers. Therefore, similar to many edge inference studies, the wireless transmission delay between the mobile user and its associated edge server can be measured or estimated by existing access network models, and the key issue considered in this paper is the cross-server transmission delay caused by distributed MoE inference among edge servers.

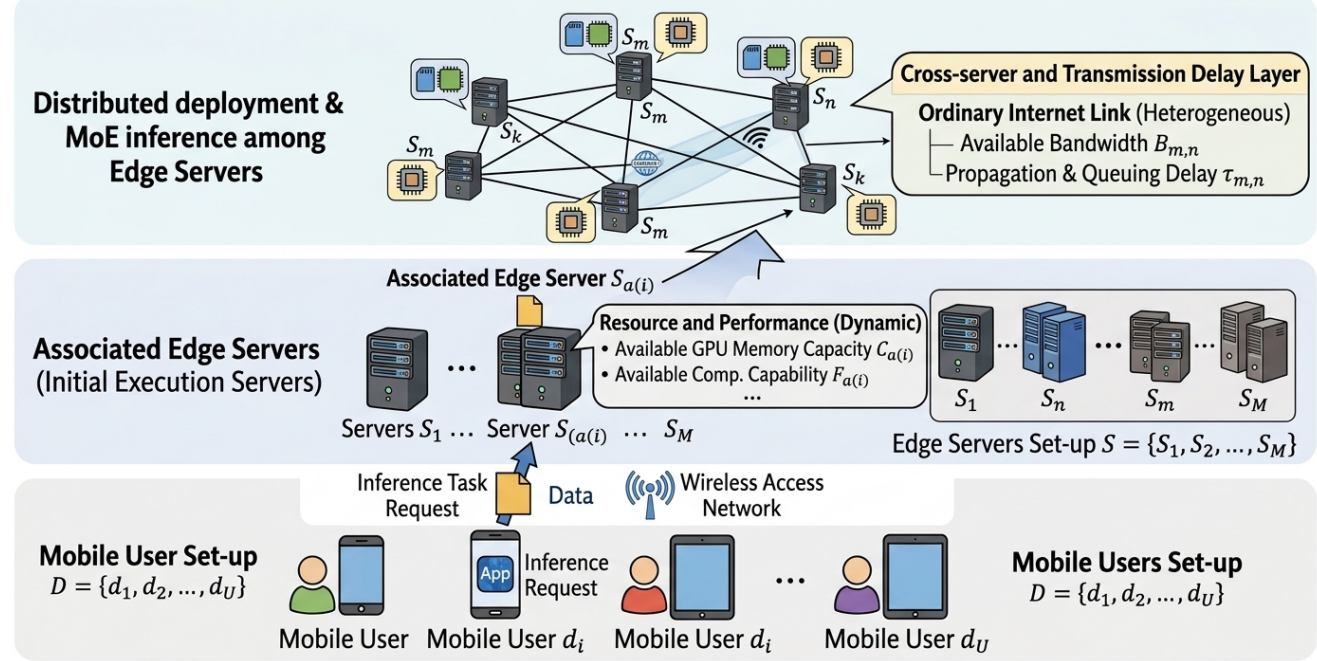

Fig.1. Network model

Since the edge servers are geographically distributed and connected through ordinary Internet links, the bandwidth among edge servers is limited. Let $B_{m,n}$ denotes the available bandwidth between edge servers $S_m$ and $S_n$, and $\tau_{m,n}$ denotes the propagation and queuing delay between them. Since the inter-server links are not dedicated, both $B_{m,n}$ and $\tau_{m,n}$ are different with different server pairs. The edge servers are also heterogeneous in computation and storage resources. Let $C_m$ denotes the available GPU memory capacity of edge server $S_m$, and $F_m$ denotes its available computation capability. In practice, $C_m$ and $F_m$ are also different with different edge servers. The workload on each edge server is dynamic, and thus the available computation resource changes during runtime.

Let the MoE model that deployed at the edge be denoted by $\mathcal{M}$. The model contains $N$ MoE layers. In the $l$-th MoE layer, the expert set is denoted by $\mathcal{E}_l = \{e_l^1, e_l^2, \dots, e_l^{E_l}\}$, where $E_l$ is the number of experts in the $l$-th MoE layer. For expert $e_l^j$, let $p_l^j$ denote its GPU memory occupation and let $c_l^j$ denote its computation workload for processing a token representation.

Since the whole MoE model is too large to be deployed on a single edge server, the experts are distributed across multiple edge servers. Let $x_{m,l,j}$ be the expert deployment decision, where $x_{m,l,j} = 1$ means expert $e_l^j$ is deployed on edge server $S_m$, otherwise, $x_{m,l,j} = 0$. Then, the memory consumption of deployed experts on edge server $S_m$ is $\sum_{l=1}^{N} \sum_{j=1}^{E_l} x_{m,l,j}\, p_l^j$. Since each edge server has limited GPU memory, the following memory constraint should be satisfied:

$$\sum_{l=1}^{N} \sum_{j=1}^{E_l} x_{m,l,j}\, p_l^j \le C_m, \quad \forall S_m \in \mathcal{S} \tag{1}$$

In addition, to guarantee that each expert can be exactly executed when required, each expert should be deployed on at least one edge server:

$$\sum_{m=1}^{M} x_{m,l,j} \geq 1, \forall l \in [1, N], j \in [1, E_l] \quad (2)$$

In traditional distributed MoE deployment on edge servers, when the target expert selected by the router is not deployed on the local server, the token representation should be transmitted to another server that stores the target expert. In this paper, we investigate a different deployment paradigm. Based on the similarity among same-layer experts, similar experts are dispersed across different edge servers, such that each edge server can preserve local substitute candidates for different expert groups. Therefore, when the target expert is not available locally, a similar local expert may be used for substitute execution to avoid cross-server token transmission.

*B. Expert Similarity and Grouping Model*

The key observation used in this paper is that experts within the same MoE layer may exhibit functional similarity. In MoE models, the router determines which experts should be activated for each token. Therefore, the routing behavior of experts over input tokens can reflect their functional preference. Additionally, the expert similarity is related to the depth of MoE layers. In general, the experts in deeper layers tend to be more similar to each other, whereas the experts in shallower layers usually have lower similarity and stronger specialization. Therefore, the expert grouping strategy should be layer-aware. For deeper layers, larger expert groups may be allowed because the experts have higher similarity. For shallower layers, more conservative grouping should be adopted to avoid grouping experts with weak substitutability. This layer-aware grouping principle is important for balancing local substitution capability and inference quality preservation.

Based on this observation, we measure the similarity between same-layer experts according to their router-induced logits representations. For the $l$-th MoE layer, let $\mathbf{X}_l = \{x_1, x_2, \dots, x_T\}$ denote the calibration token representations input to the router, where $T$ is the number of calibration tokens. Let $\mathbf{W}_r^l$ denote the routing weight matrix of the $l$-th MoE layer. Then, the routing logits matrix can be calculated as:

$$\mathbf{H}_l = \mathbf{W}_r^l \mathbf{X}_l^T \quad (3)$$

The $j$-th row vector of $\mathbf{H}_l$, denoted by $\mathbf{H}_{l,j,*}$, represents the routing response of expert $e_l^j$ over the calibration tokens. Therefore, the similarity between experts $e_l^i$ and $e_l^j$ in the same layer can be calculated by the cosine similarity of their logits row vectors:

$$\text{Sim}_l(i,j) = \frac{\mathbf{H}_{l,i,*} \cdot \mathbf{H}_{l,j,*}}{\|\mathbf{H}_{l,i,*}\| \|\mathbf{H}_{l,j,*}\|} \quad (4)$$

A larger $\text{Sim}_l(i,j)$ indicates that these two experts have more similar routing responses and are more likely to process similar token patterns. Therefore, they can be considered as potential substitute experts for each other.

Besides expert similarity, the activation frequency of each expert is also considered. Let $n_l^j$ denote the number of times that expert $e_l^j$ is activated on the calibration dataset. Then, the activation frequency of expert $e_l^j$ can be calculated as:

$$f_l^j = \frac{n_l^j}{\sum_{k=1}^{E_l} n_l^k} \quad (5)$$

Experts with high activation frequencies are more frequently requested during inference. In this paper, these frequently activated experts are regarded as dominant experts, and they are used as the centers of expert groups.

Let $\mathcal{G}_l = \{G_l^1, G_l^2, \dots, G_l^{K_l}\}$ denote the expert grouping result of the $l$-th MoE layer, where $K_l$ is the number of expert groups in this layer. Each group contains one dominant expert and several non-dominant experts that are similar to it. If expert $e_l^j$ is a non-dominant expert, it is assigned to the group whose dominant expert has the largest similarity with it. In this way, the experts in the same group have similar routing behavior and can provide substitute candidates for each other. Since the similarity distribution is layer-dependent, both the number of groups $K_l$ and the similarity threshold $\theta_l$ are made layer-aware. We propose a depth-dependent threshold:

$$\theta_l = \theta_{min} + (\theta_{max} - \theta_{min}) \cdot \frac{N-l}{N-1} \quad (6)$$

The (6) means that the shallow layers use a stricter threshold (closer to $\theta_{\max}$) to preserve specialization, while deeper layers use a looser threshold (closer to $\theta_{\min}$) to allow larger substitutable groups. The detailed determination of $K_l$ and the grouping procedure are described in Section IV.

For the target expert $e_l^r$, the feasible substitute expert set is defined as:

$$\mathcal{R}_{l,r} = \{e_l^k \mid e_l^k \in G_l^g, e_l^r \in G_l^g, \text{Sim}_l(r,k) \geq \theta_l\} \quad (7)$$

where $\theta_l$ is the similarity threshold of the $l$-th MoE layer. The feasible execution expert set of target expert $e_l^r$ is defined as:

$$\mathcal{V}_{l,r} = \{e_l^r\} \cup \mathcal{R}_{l,r} \quad (8)$$

In (8), $e_l^r$ represents exact execution, while the experts in $\mathcal{R}_{l,r}$ represent feasible substitute execution.

The expert grouping result provides candidate substitute experts for each target expert. When the target expert is not selected for execution, a similar expert in the same group may be used for substitute execution. Since the substitute expert is different from the target expert, such execution may cause inference quality loss. Therefore, the quality loss caused by expert substitution should be considered in the inference.

*C. Similarity-Aware Inference Model*

For an inference request of user $d_i$, let $Q_i$ denote its token set. For token $q \in Q_i$, suppose that the target expert selected by the router in the $l$-th MoE layer is denoted by $e_l^{r_l(q)}$, where $r_l(q) \in \{1,2,\dots,E_l\}$. In traditional distributed MoE inference, when the target expert selected by the router is not deployed on the current edge server, the token representation is transmitted to another edge server that stores the target expert. After the expert computation is completed, the computed representation is returned to the access edge server for the following layers. However, this fixed-return strategy may introduce unnecessary cross-server transmission. For example, if the token is transmitted to a remote server for the current MoE layer and a feasible expert for the next MoE layer is also deployed on the same server, returning the intermediate representation to the access edge server will cause redundant transmission.

Based on the above expert similarity and grouping model, this paper investigates a token trajectory aware inference paradigm. The token representation is not forced to return to the access edge server after remote expert execution. Instead, for each token in each MoE layer, the system jointly determines the execution server and the actually executed expert. In this way, the token can be processed by the target expert or by a similar substitute expert on a suitable edge server. This strategy can reduce cross-server transmission while controlling the quality loss caused by substitution.

Since the router module is much smaller than the expert modules, we assume that the routers are replicated on all edge servers. Therefore, the routing decision of each MoE layer can be obtained at the server where the token representation is located. The non-expert modules are assumed to be replicated on all edge servers. In this paper, we mainly focus on the placement and invocation of MoE experts, which dominate the memory occupation and cross-server transmission overhead. Let $u_{i,q,l,m,k} = 1$, if token $q$ of user $d_i$ is processed by expert $e_l^k$ on server $S_m$ ; otherwise, $u_{i,q,l,m,k} = 0$ . The variable $u_{i,q,l,m,k}$ represents both the execution server and the executed expert. If $k = r_l(q)$, the target expert is selected and exact execution is adopted. If $k \neq r_l(q)$, a substitute expert is selected. In this case, the selected expert should be in the feasible substitute expert set of the target expert. Therefore, for each token in each MoE layer, only one server-expert pair is selected, which is expressed as:

$$\sum_{m=1}^{M} \sum_{k=1}^{E_l} u_{i,q,l,m,k} = 1, \forall d_i \in \mathcal{D}, q \in \mathcal{Q}_i, l \in [1, N] \quad (9)$$

Moreover, token $q$ can be processed by expert $e_l^k$ on server $S_m$ only when this expert is deployed on $S_m$. Therefore, we have:

$$u_{i,q,l,m,k} \le x_{m,l,k}, \forall d_i, q, l, m, k \quad (10)$$

The selected expert should be either the target expert or a feasible substitute expert. Therefore,

$$u_{i,q,l,m,k} = 0, if\ e_l^k \notin \mathcal{V}_{l,r_l(q)} \quad (11)$$

For describing the execution server of each token in each MoE layer, we define:

$$\mu_{i,q,l,m} = \sum_{k=1}^{E_l} u_{i,q,l,m,k} \quad (12)$$

If $\mu_{i,q,l,m} = 1$, token $q$ of user $d_i$ is executed on server $S_m$ in the $l$-th MoE layer. At the beginning of inference, the token representation is located at the access edge server $S_{a(i)}$. After the token is processed on server $S_m$ in the $l$-th MoE layer, the output representation stays on $S_m$ and becomes the input representation of the next MoE layer. Therefore, the token trajectory across different MoE layers can be described by the execution server sequence $(S_{m_1}, S_{m_2}, \dots, S_{m_N})$, where $\mu_{i,q,l,m_l} = 1$. If $m_l = m_{l+1}$, no cross-server transmission is required between the $l$-th and $(l+1)$-th MoE layers. If $m_l \neq m_{l+1}$ , the intermediate token representation should be transmitted from $S_{m_l}$ to $S_{m_{l+1}}$.

In this model, whether the token representation returns to the access edge server is not an independent decision. It is determined by the execution server selected in the next MoE layer. If the next-layer execution server is the access edge server, the token returns to $S_{a(i)}$. Otherwise, it can stay on the current server or migrate to another server.

*D. Inference Delay Model*

The inference delay of each token in each MoE layer consists of two parts: transmission delay and computation delay. Let $z_l$ denote the size of the intermediate token representation that needs to be transmitted before the $l$-th MoE layer when two adjacent execution servers are different. If the input and output representation sizes of different MoE layers are different, $z_l$ can be set as the corresponding layer-specific data size.

For the first MoE layer, the token representation is initially located at the access edge server $S_{a(i)}$ . If the first-layer execution server is not the access edge server, the token representation should be transmitted from $S_{a(i)}$ to the selected execution server. Therefore, the transmission delay of the first MoE layer is:

$$T_{i,q,1}^{trans} = \sum_{m=1}^{M} \mu_{i,q,1,m} \mathbf{1}(m \neq a(i)) \left( \frac{z_1}{B_{a(i),m}} + \tau_{a(i),m} \right) \quad (13)$$

where $\mathbf{1}(\cdot)$ is the indicator function. For the $l$-th MoE layer with $l \ge 2$, the token representation is located at the execution server of the $(l-1)$-th MoE layer before entering the $l$-th MoE layer. Therefore, the transmission delay before the $l$-th MoE layer is:

$$T_{i,q,l}^{trans} = \sum_{m=1}^{M} \sum_{n=1}^{M} \mu_{i,q,l-1,m}\, \mu_{i,q,l,n} \mathbf{1}(m \neq n) \left( \frac{z_l}{B_{m,n}} + \tau_{m,n} \right), l \ge 2 \quad (14)$$

According to (14), if two adjacent MoE layers of the same token are executed on the same edge server, the corresponding transmission delay is zero.

The computation delay of token $q$ in the $l$-th MoE layer is determined by the selected expert and the execution server. Therefore, the computation delay is calculated as:

$$T_{i,q,l}^{comp} = \sum_{m=1}^{M} \sum_{k=1}^{E_l} u_{i,q,l,m,k} \cdot \frac{c_l^k}{F_m} \quad (15)$$

Then, the total delay of token $q$ in the $l$-th MoE layer is:

$$T_{i,q,l} = T_{i,q,l}^{trans} + T_{i,q,l}^{comp} \quad (16)$$

The total inference delay of user $d_i$ is:

$$T_i = \sum_{q \in \mathcal{Q}_i} \sum_{l=1}^{N} T_{i,q,l} \quad (17)$$

*E. Substitution Quality Model*

Although local substitute execution can reduce cross-server transmission delay, it may introduce inference quality loss because the substitute expert is not exactly the same as the target expert. Therefore, the quality loss caused by substitution should be considered in the problem formulation.

For token $q$ in the $l$-th MoE layer, the target expert is $e_l^{r_l(q)}$. If the actually selected expert is $e_l^k$, the substitution quality loss is defined as:

$$\mathcal{Q}_l(r_l(q), k) = \begin{cases} 0, & k = r_l(q), \\ \phi\left( \frac{1 - \mathrm{Sim}_{\,l}(r_l(q), k)}{2} \right), & k \neq r_l(q), \end{cases} \quad (18)$$

where $\phi(\cdot)$ is a non-decreasing function. Since $\mathrm{Sim}_{\,l}(r_l(q), k)$ is calculated by cosine similarity, the term $\frac{1 - \mathrm{Sim}_{\,l}(r_l(q), k)}{2}$ is used to normalize the similarity distance into $[0,1]$. If the selected expert is the target expert, the substitution quality loss is zero. If the selected expert is a substitute expert, the quality loss is related to the similarity between the target expert and the selected substitute expert. Then, the substitution quality loss of token $q$ in the $l$-th MoE layer is:

$$Q_{i,q,l} = \sum_{m=1}^{M} \sum_{k=1}^{E_l} u_{i,q,l,m,k} \cdot \mathcal{Q}_l(r_l(q), k) \quad (19)$$

The total substitution quality loss of user $d_i$ is calculated as:

$$Q_i = \sum_{q \in \mathcal{Q}_i} \sum_{l=1}^{N} Q_{i,q,l} \quad (20)$$

To guarantee the inference quality, the total substitution quality loss of each user should not exceed the maximum acceptable value, which is expressed as:

$$Q_i \le Q_i^{max}, \forall d_i \in \mathcal{D} \quad (21)$$

The above model indicates that the server selection and expert selection are related to both inference delay and substitution quality loss. For example, if the target expert of the next MoE layer is not deployed on the current server, the token can either be processed by a similar substitute expert on the current server or be transmitted to another server that stores the target expert or a better substitute expert. The former may reduce transmission delay but introduce substitution quality loss, while the latter may reduce quality loss but increase transmission delay. Therefore, it is necessary to jointly consider inference delay and substitution quality loss when determining the token trajectory and expert selection strategy.

*F. Problem Statement*

From the above subsections, we obtain the inference delay $T_i$ and substitution quality loss $Q_i$ of user $d_i$. The purpose of this paper is to minimize the inference delay and substitution quality loss by jointly optimizing the expert grouping strategy, expert deployment strategy, and token-trajectory-aware server-expert selection strategy. The optimization variables include the expert grouping strategy $\mathcal{G} = \{\mathcal{G}_1, \mathcal{G}_2, \ldots, \mathcal{G}_N\}$, the expert deployment strategy $\mathbf{X} = \{x_{m,l,j} \mid S_m \in \mathcal{S}, l \in [1,N], j \in [1,E_l]\}$, and the server-expert selection strategy $\mathbf{U} = \{u_{i,q,l,m,k} \mid d_i \in \mathcal{D}, q \in \mathcal{Q}_i, l \in [1,N], S_m \in \mathcal{S}, k \in [1,E_l]\}$. Since the inter-server links and the computation resources are shared, we further bound the per-server load within each scheduling slot. Let $\mathcal{D}(t)$ denote the set of requests served in slot $t$. The computation workload assigned to edge server $S_m$ in slot $t$ is:

$$W_m(\mathbf{X},\mathbf{U};t) = \sum_{d_i \in \mathcal{D}(t)} \sum_{q \in \mathcal{Q}_i} \sum_{l=1}^{N} \sum_{k=1}^{E_l} u_{i,q,l,m,k} \cdot c_l^k \quad (22)$$

Then, the problem P0 can be expressed as:

$$\mathbf{P0}: \min \{\textstyle\sum_{i=1}^{U} T_i, \sum_{i=1}^{U} Q_i\}$$

s.t.

$$\sum_{l=1}^{N} \sum_{j=1}^{E_l} x_{m,l,j}\, p_l^j \le C_m \quad \text{(c.1)}$$

$$\sum_{m=1}^{M} x_{m,l,j} \ge 1, \forall l \in [1,N], j \in [1,E_l] \quad \text{(c.2)}$$

$$x_{m,l,j} \in \{0,1\}, \forall S_m \in \mathcal{S}, l \in [1,N], j \in [1,E_l] \quad \text{(c.3)}$$

$$u_{i,q,l,m,k} \in \{0,1\}, \forall d_i, q, l, m, k \quad \text{(c.4)}$$

$$\sum_{m=1}^{M} \sum_{k=1}^{E_l} u_{i,q,l,m,k} = 1, \forall d_i \in \mathcal{D}, q \in \mathcal{Q}_i, l \in [1,N] \quad \text{(c.5)}$$

$$u_{i,q,l,m,k} \le x_{m,l,k}, \forall d_i, q, l, m, k \quad \text{(c.6)}$$

$$u_{i,q,l,m,k} = 0, \text{if } e_l^k \notin \mathcal{V}_{l,r_l(q)} \quad \text{(c.7)}$$

$$Q_i \le Q_i^{max}, \forall d_i \in \mathcal{D} \quad \text{(c.8)}$$

$$W_m(\mathbf{X},\mathbf{U};t) \le F_m \Delta t, \forall S_m \in \mathcal{S}, \forall t \quad \text{(c.9)}$$

In P0, $T_i$ is the total inference delay of user $d_i$, and $Q_i$ is the total substitution quality loss of user $d_i$. Constraint (C.1) means that the memory consumption of the deployed experts on each edge server should not exceed its GPU memory capacity. Constraint (C.2) guarantees that each expert is deployed on at least one edge server. Constraint (C.3) means that the expert deployment decision is binary. Constraint (C.4) means that the server-expert selection decision is binary. Constraint (C.5) means that each token in each MoE layer should be processed by one server-expert pair. Constraint (C.6) means that an expert can be selected on a server only when it is deployed on that server. Constraint (C.7) guarantees that the selected expert is either the target expert or a feasible substitute expert. Constraint (C.8) means that the substitution quality loss of each user should not exceed the maximum acceptable value. Constraint (C.9) means the workload assigned to each edge server within a scheduling slot $\Delta t$ should not exceed its available computation capability.

This problem is difficult to solve for the following reasons. First, the optimization objectives are opposite to each other. Selecting a nearby substitute expert may reduce transmission delay, but it may increase substitution quality loss. Selecting the target expert can avoid substitution quality loss, but it may introduce additional cross-server transmission delay. Second, the expert grouping strategy, expert deployment strategy, and server-expert selection strategy are discrete decisions, which makes the solution space very large. Third, the execution server selected in one MoE layer determines the source server of the next-layer transmission. Therefore, the server-expert selection decisions of different layers are coupled with each other. Fourth, the edge servers are heterogeneous, and the inter-server network conditions are different. Therefore, finding the optimal solution of P0 is difficult.

Since the objectives in P0 are different, a weight-based approach is introduced to construct the utility function. Before constructing the utility function, $T_i$ and $Q_i$ are normalized. Let $\tilde{T}_i = T_i / T_i^{ref}$ and $\tilde{Q}_i = Q_i / Q_i^{max}$ denote the normalized inference delay and normalized substitution quality loss of user $d_i$, respectively. Moreover, $T_i^{ref}$ denotes the inference delay of user $d_i$ under the pure exact-execution strategy on the same deployment, and $Q_i^{max}$ is the quality budget of user $d_i$. Then, the utility function is expressed as:

$$U(\mathcal{G},\mathbf{X},\mathbf{U}) = \omega_T \sum_{i=1}^{U} \tilde{T}_i + \omega_Q \sum_{i=1}^{U} \tilde{Q}_i \quad (23)$$

where $\omega_T$ and $\omega_Q$ are the weights of inference delay and substitution quality loss, respectively, and $\omega_T + \omega_Q = 1$.

Thus, P0 can be transformed into the following single-objective optimization problem:

$$\mathbf{P1}: \min_{\mathcal{G},\mathbf{X},\mathbf{U}} U(\mathcal{G},\mathbf{X},\mathbf{U})$$

s.t. constraints (C.1) to (C.9).

According to (23), the weights can be adjusted according to different service requirements. If low latency is more important, a larger $\omega_T$ can be used. If inference quality is more important, a larger $\omega_Q$ can be adopted. Therefore, the weight-based approach can flexibly achieve the tradeoff between inference delay and substitution quality loss.

## IV. Similarity- aware Expert Grouping and Distributed Deployment

In Section III, the distributed edge MoE inference problem has been formulated. Since P1 contains expert grouping, expert deployment, and token-level server-expert selection, it is difficult to solve it directly. In this section, we propose a similarity-aware expert grouping and distributed deployment strategy to obtain the expert grouping result $\mathcal{G}$ and the deployment strategy $\mathbf{X}$. Based on the deployment result, the runtime token trajectory and server-expert selection strategy will be introduced in Section V.

As shown in Fig.2, the proposed strategy contains two main parts: 1) layer-aware expert grouping based on router-induced similarity and activation frequency, and 2) representative coverage deployment across heterogeneous edge servers. The purpose is to increase the probability that each token can find the routed expert or a feasible substitute expert on a proper edge server, thereby reducing unnecessary cross-server transmission during inference.

*A. Layer-Aware Expert Grouping*

The purpose of expert grouping is to put functionally similar experts of the same MoE layer into the same group, so that each group can provide substitute candidates for its members. Based on the expert similarity in (4) and the activation frequency in (5), we propose a layer-aware grouping strategy. As discussed in Section III, the experts in deeper layers tend to be more similar, while the experts in shallower layers are more specialized. Therefore, instead of fixing the number of groups, we determine the groups according to the layer-aware similarity threshold $\theta_l$ in (6). In this way, shallow layers naturally produce more groups with stronger specialization, while deep layers produce fewer and larger groups with stronger substitutability.

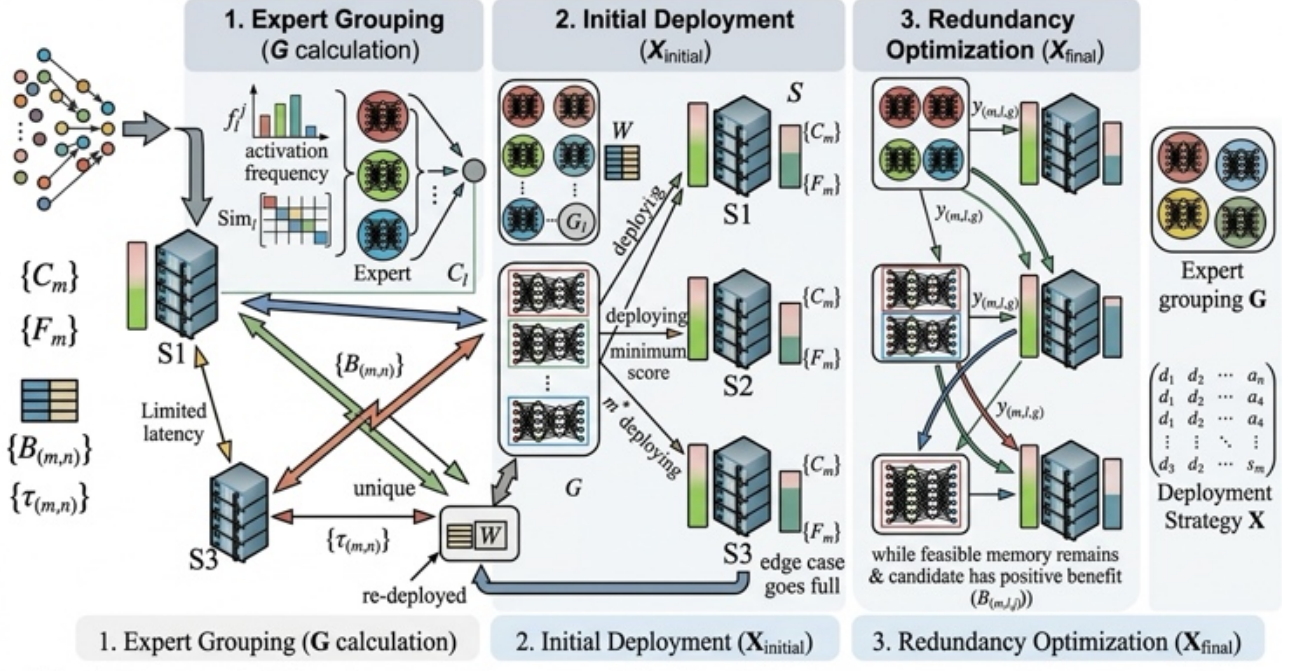

Fig.2. Similarity-aware expert grouping and distributed deployment strategy

The grouping is performed by a greedy representative-cover procedure. Let $\mathcal{C}_l$ denote the set of dominant experts of the $l$-th MoE layer, which is initialized as empty. The experts are processed in the order of decreasing activation frequency, so that frequently activated experts are preferentially selected as dominant experts. For expert $e_l^j$, it becomes a new dominant expert if it is not similar enough to any existing dominant expert:

$$e_l^j \in \mathcal{C}_l \text{if} \max_{e_l^c \in \mathcal{C}_l} \text{Sim}_l(j, c) < \theta_l \tag{24}$$

Otherwise, $e_l^j$ is regarded as a non-dominant expert and assigned to the group whose dominant expert has the largest similarity with it:

$$g(j) = \arg \max_{e_l^c \in \mathcal{C}_l} \text{Sim}_l(j, c) \tag{25}$$

Therefore, the $g$-th group of the $l$-th MoE layer is defined as:

$$G_l^g = \{e_l^{c_g}\} \cup \{ e_l^j \mid g(j) = c_g \} \tag{26}$$

where $e_l^{c_g}$ is the dominant expert of group $G_l^g$. The number of groups in the $l$-th MoE layer is then $K_l = \mid \mathcal{C}_l \mid$, which is determined automatically by the layer-aware threshold $\theta_l$ rather than set manually. In this way, the experts in the same group have similar routing behavior and can serve as feasible substitute experts for each other, while the experts assigned to different groups remain functionally distinct.

*B. Importance-Aware Expert Deployment*

After grouping, the experts should be deployed across the edge servers under the memory constraint. Two principles are considered. First, important experts should be deployed with priority. Second, the experts of the same group should be dispersed across different edge servers, so that each edge server can locally cover as many groups as possible and provide more local substitute candidates.

An expert is regarded as important if it is frequently activated and if it can well represent the other experts in its group. Therefore, the representativeness of expert $e_l^j$ is defined as the average similarity to the experts in its group:

$$\rho_l^j = \frac{1}{|G_l^{g(j)}|} \sum_{e_l^k \in G_l^{g(j)}} \text{Sim}_l\,(j, k) \tag{27}$$

Then, the importance of expert $e_l^j$ is defined as:

$$I_l^j = \alpha_f\, f_l^j + \alpha_r\, \rho_l^j \tag{28}$$

where $\alpha_f$ and $\alpha_r$ are non-negative weights with $\alpha_f + \alpha_r = 1$. A larger $I_l^j$ means that expert $e_l^j$ is activated more frequently and can substitute its group members with smaller quality loss.

The deployment guarantees that each expert is deployed at least once, as required by (c.2). Let $b_m$ denote the currently used memory of edge server $S_m$, which is initialized as zero. For expert $e_l^j$, the memory-feasible server set is defined as:

$$\mathcal{F}_{l,j} = \{ S_m \in \mathcal{S} \mid b_m + p_l^j \le C_m \} \tag{29}$$

To disperse the same-group experts and balance the memory load, a deployment score is defined for placing expert $e_l^j$ on edge server $S_m$:

$$\text{Score}(m, l, j) = \lambda_1 \cdot \frac{b_m}{C_m} + \lambda_2 \cdot \eta_{m,l,g(j)} \tag{30}$$

where $\eta_{m,l,g(j)} = \sum_{e_l^k \in G_l^{g(j)}} x_{m,l,k}$ is the number of experts of the same group already deployed on $S_m$, and $\lambda_1 + \lambda_2 = 1$. The first term prefers a less loaded server, and the second term penalizes placing experts of the same group on the same server, so that similar experts are spread over different servers. Therefore, the expert is deployed to the feasible server with the minimum deployment score:

$$m^* = \arg \min_{S_m \in \mathcal{F}_{l,j}} \text{Score}(m, l, j) \tag{31}$$

Then $x_{m^*,l,j} = 1$ is set, and the used memory is updated as $b_{m^*} \leftarrow b_{m^*} + p_l^j$.

It is worth noting that the feasible server set $\mathcal{F}_{l,j}$ may be temporarily empty due to memory fragmentation. To handle this case, the mandatory deployment processes the experts of each layer in the order of non-increasing memory demand $p_l^j$, which is a first-fit-decreasing manner that reduces fragmentation. If $\mathcal{F}_{l,j}$ is still empty when expert $e_l^j$ is processed, the expert is pushed into a waiting list. After the current layer is processed, the experts in the waiting list are re-deployed to the server with the largest remaining memory. As long as the total available memory of the edge servers is sufficient to store all experts at least once, this procedure can deploy every expert at least once, so that constraints (c.1) to (c.3) are satisfied.

*C. Importance-Aware Redundant Deployment*

After each expert is deployed at least once, the remaining memory of edge servers can be used to deploy redundant replicas. The purpose of redundant deployment is to improve the local coverage of important groups, so that more tokens can find an exact expert or a feasible substitute expert near their current location. To describe the local coverage, an indicator is defined for whether edge server $S_m$ can locally serve group $G_l^g$:

$$y_{m,l,g} = \begin{cases} 1, & \sum_{e_l^k \in G_l^{g(j)}} x_{m,l,i} \ge 1, \\ 0, & \text{otherwise.} \end{cases} \tag{32}$$

If $y_{m,l,g} = 1$, edge server $S_m$ already hosts at least one expert of group $G_l^g$, and a token routed to this group can be locally executed on $S_m$ by exact or substitute execution. Therefore, deploying a replica is more beneficial when the target group is not yet locally covered. For a candidate replica of expert $e_l^j$ on edge server $S_m$, the redundancy benefit is defined as:

$$B(m,l,j) = \frac{\pi_m \cdot I_l^j \cdot (1-y_{m,l,g(j)})}{p_l^j} \quad (33)$$

where $\pi_m$ denotes the relative amount of tokens whose trajectory reaches $S_m$, which is estimated from the user-server association and the calibration access distribution. The numerator means that a replica is more valuable when it serves a busy server, carries a high-importance expert, and covers a group that is not yet locally available. The denominator $p_l^j$ normalizes the benefit by the memory cost, so that the replica with the highest benefit per unit memory is preferred.

Let $x_{m,l,j}^{\text{red}}$ denote the redundant deployment decision. The redundant deployment is formulated as:

$$\max_{\{x_{m,l,j}^{\text{red}}\}} \sum_{m=1}^{M} \sum_{l=1}^{N} \sum_{j=1}^{E_l} x_{m,l,j}^{\text{red}} \quad (34)$$

subject to the memory constraint of each edge server:

$$b_m + \sum_{l=1}^{N} \sum_{j=1}^{E_l} x_{m,l,j}^{\text{red}} p_l^j \leq C_m, \forall S_m \in \mathcal{S} \quad (35)$$

Since (34) and (35) are knapsack-like problem over the remaining memory, it is solved greedily. In each iteration, the candidate replica with the largest redundancy benefit that still fits the remaining memory is selected and deployed. Then the coverage indicator $y_{m,l,g}$, the used memory $b_m$, and the benefit of the related candidates are updated. The iteration stops when no replica can be deployed or no candidate provides a positive benefit. This greedy rule preferentially extends the local coverage of important groups on busy servers, which directly increases the probability of local exact execution and local substitute execution during runtime.

*D. Grouping and Deployment Algorithm*

Based on the above models, as shown in Table I, the similarity-aware expert grouping and distributed deployment procedure is summarized in Algorithm 1.

Table I

Alaogirtim1: similarity aware expert grouping and deployment

**Input:** MoE model $\mathcal{M}$, edge server set $\mathcal{S}$, calibration token set, memory capacity $\{C_m\}$, computation capacity $\{F_m\}$, bandwidth $\{B_{m,n}\}$, latency $\{\tau_{m,n}\}$.
**Output:** Expert grouping result $\mathcal{G}$ and deployment strategy $\boldsymbol{X}$.
1. Initialize $X=0$, $\mathcal{G}=\emptyset$, and $b_m=0$ for each $S_m \in \mathcal{S}$;
2. for each MoE layer $l=1$ to $N$ do
3. Calculate routing logits matrix $H_l$ on the calibration token set by (3)
4. Compute pairwise expert similarity matrix $\text{Sim}_l$ by (4)
5. Compute activation frequency $f_l^j$ by (5); sort experts in decreasing $f_l^j$
6. Initialize center set $\mathcal{C}_l=\emptyset$
7. for each expert $e_l^j$ in the sorted order do
8. if $\max_{e_l^c \in \mathcal{C}_l} \text{Sim}_l(j,c) < \theta_l$ then add $e_l^j$ into $\mathcal{C}_l$ as a new center by (24)
9. else assign $e_l^j$ to the group of its most similar center by (25)
10. end if
11. end for
12. Compute representativeness $\rho_l^j$ by (27) and importance $I_l^j$ by (28)
13. end for
14. for each MoE layer $l=1$ to $N$ do
15. Sort experts in non-increasing memory demand $p_l^j$; set waiting list $\mathcal{W}=\emptyset$
16. for each expert $e_l^j$ in the sorted order do
17. Find feasible server set $\mathcal{F}_{l,j}$ by (29)
18. if $\mathcal{F}_{l,j} \neq \emptyset$ then
19. Deploy $e_l^j$ to server $m^*$ with the minimum score by (30) to (31); update $b_{m^*}$
20. else push $e_l^j$ into $\mathcal{W}$
21. end if
22. end for
23. Re-deploy each expert in $\mathcal{W}$ to the server with the largest remaining memory
24. end for
25. Compute redundancy benefit $B(m,l,j)$ for all feasible server-expert pairs by (33)
26. while feasible memory remains and some candidate has positive benefit do
27. Deploy the replica with the largest benefit; update $y_{m,l,g}$, $b_m$, and the related benefits
28. end while
29. return $\mathcal{G}$ and $X$

The algorithm contains three parts.

**Line1-Line13:** Calculating expert similarity and constructing layer-aware expert groups. After initialization, for each MoE layer the routing logits matrix is first calculated using the calibration token set, and the pairwise expert similarity matrix is then obtained. The activation frequencies are calculated, and the experts are sorted in decreasing order, so that frequently activated experts are preferentially examined. Each expert is processed in turn: if its largest similarity to the existing group centers is smaller than the layer-dependent threshold $\theta_l$, it becomes a new group center; otherwise, it is assigned to the group whose center has the largest similarity with it. In this way, the number of groups is determined automatically by the layer-aware threshold rather than set manually. After the groups are formed, the representativeness and the importance of each expert are computed.

**Line14-Line24:** Performing representative coverage deployment. For each MoE layer, the experts are first sorted in non-increasing memory demand to reduce memory fragmentation. Then, each expert is deployed to the feasible edge server with the minimum deployment score, which jointly considers the memory occupation and the same-group dispersion, so that similar experts are spread over different servers. If no feasible server is currently available, the expert is pushed into a waiting list and re-deployed to the server with the largest remaining memory after the current layer is processed. This step guarantees that each expert is deployed at least once and that constraints (c.1) to (c.3) are satisfied.

**Line25-Line29:** Performing importance-aware redundant deployment. After the initial deployment, the algorithm calculates the redundancy benefit of each feasible server-expert pair, which favors the replica that covers an uncovered group of an important expert on a busy server with the highest benefit per unit memory. Then, additional replicas are deployed greedily, and the local coverage indicator, the used memory, and the related benefits are updated after each deployment. The iteration stops when no feasible memory resource is left or no candidate provides a positive benefit. Finally, the grouping result $\mathcal{G}$ and the deployment strategy $\boldsymbol{X}$ are returned.

The computational complexity of Algorithm 1 is analyzed as follows. For the $l$-th MoE layer, calculating the similarity matrix requires $O(E_l^2 T)$ operations, where $T$ is the number of calibration tokens. The grouping step requires $O(E_l K_l)$ operations. The mandatory deployment step requires $O(M \sum_{l=1}^{N} E_l)$ operations. If the number of redundant deployment iterations is $R_{\text{rep}}$, the redundant deployment step requires $O(R_{\text{rep}}\, M \sum_{l=1}^{N} E_l)$ operations. Therefore, the total complexity of Algorithm 1 is: $O\left(\sum_{l=1}^{N} E_l^2 T + \sum_{l=1}^{N} E_l K_l + M \sum_{l=1}^{N} E_l + R_{\text{rep}}\, M \sum_{l=1}^{N} E_l\right)$, which is polynomial in the number of experts, edge servers, and calibration tokens. Therefore, the offline grouping and deployment can be completed before runtime, and the obtained $\mathcal{G}$ and $\boldsymbol{X}$ are used by the online server-expert selection strategy in Section V.

## V. Token Trajectory Aware Online Server-expert Selection

In Section IV, the offline grouping result $\mathcal{G}$ and the deployment strategy $\boldsymbol{X}$ have been obtained. Based on these results, the remaining task is to determine the server-expert selection strategy $\boldsymbol{U}$ at runtime. As discussed in Section III, the execution server selected in one MoE layer determines the source server of the next-layer transmission, so the per-layer decisions of the same token are coupled with each other. Therefore, a myopic strategy that only minimizes the cost of the current MoE layer may move the token to a server that is unfavorable for the following layers, which introduces extra cross-server transmission. To address this issue, we propose a token-trajectory-aware online server-expert selection strategy, as shown in Fig.3. For each token in each MoE layer, the strategy jointly determines the execution server and the executed expert by considering both the current-layer cost and an estimated look-ahead cost of the following layers, while respecting the per-user quality budget.

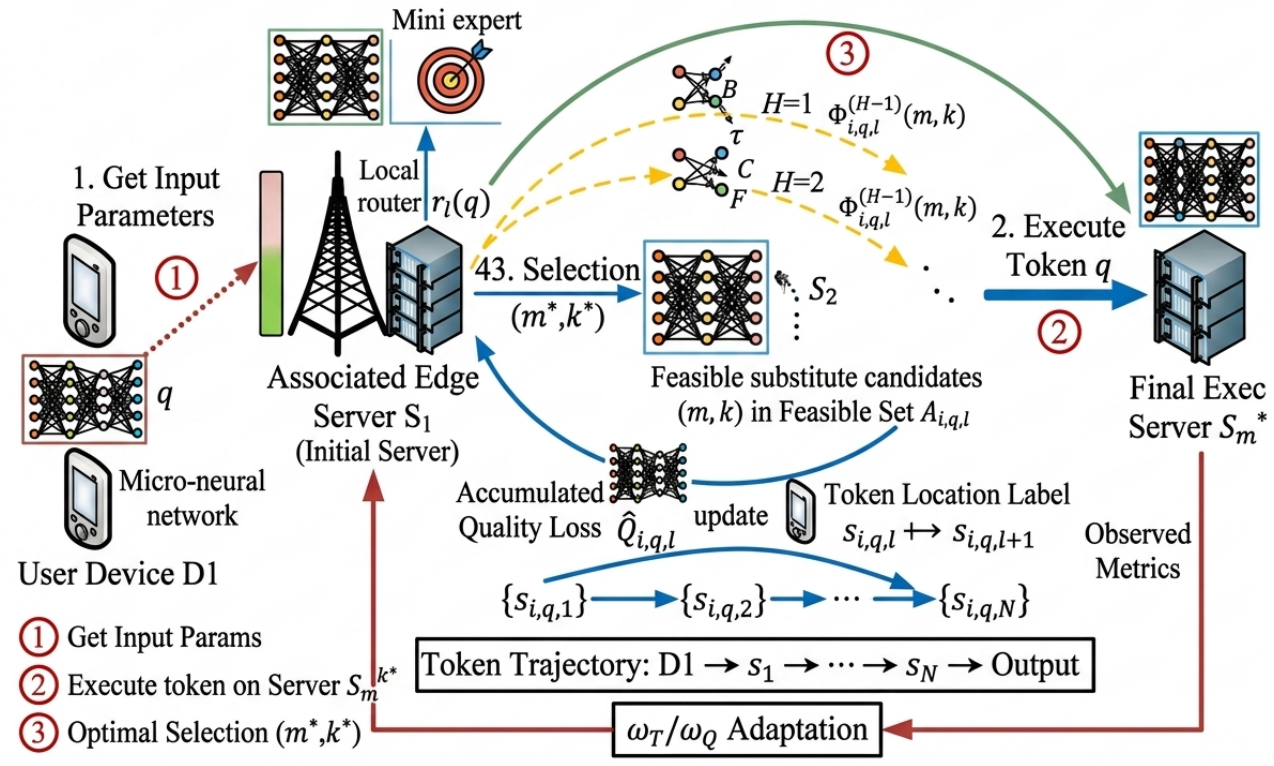

Fig.3. Token-trajectory-aware online server-expert selection strategy

### A. Online Decision Model and Per-Step Cost

The online selection processes the tokens layer by layer. For token $q$ of user $d_i$, let $s_{i,q,l}$ denote the server where the token representation resides before entering the $l$-th MoE layer. According to the token-trajectory model in Section III, the initial location is the access edge server, and the location of each subsequent layer is the execution server of the previous layer:

$$s_{i,q,1} = a(i),\, s_{i,q,l} = \arg_m(\mu_{i,q,l-1,m} = 1),\, l \geq 2 \quad (36)$$

When the token is processed by expert $e_l^k$ on server $S_m$ in the $l$-th MoE layer, the per-step delay consists of the transmission delay from the current location $s_{i,q,l}$ to $S_m$ and the computation delay on $S_m$:

$$d_{i,q,l}(m,k) = \mathbb{1}(s_{i,q,l} \neq m)\left(\frac{z_l}{B_{s_{i,q,l},m}} + \tau_{s_{i,q,l},m}\right) + \frac{c_l^k}{F_m} \quad (37)$$

If the selected server is the current location, the transmission term vanishes and only the computation delay remains. Following the normalized weighting in (23), the per-step cost of selecting the server-expert pair $(m, k)$ is defined as:

$$g_{i,q,l}(m,k) = \omega_T \cdot \frac{d_{i,q,l}(m,k)}{d_i^{\text{ref}}} + \omega_Q \cdot \frac{Q_l(r_l(q),k)}{\bar{Q}_i} \quad (38)$$

where $d_i^{\text{ref}}$ is the reference per-step delay of user $d_i$ under pure local exact execution, and $\bar{Q}_i$ is the per-token quality budget defined below. The two terms are dimensionless and share the same weights as the global utility, so that the online step cost is consistent with the offline objective.

### B. Quality-Budget-Aware Feasible Candidate Set

To guarantee the per-user quality constraint (c.8) in an online manner, the user-level budget $Q_i^{\max}$ is evenly allocated to the tokens of user $d_i$:

$$\bar{Q}_i = \frac{Q_i^{\max}}{|Q_i|} \quad (39)$$

Let $\hat{Q}_{i,q,l}$ denote the accumulated substitution quality loss of token $q$ before the $l$-th MoE layer, which is initialized as zero. Then, the runtime feasible candidate set of token $q$ in the $l$-th MoE layer is defined as:

$$\mathcal{A}_{i,q,l} = \{(m,k) \mid x_{m,l,k} = 1, e_l^k \in V_{l,r_l(q)}, \hat{Q}_{i,q,l} + Q_l(r_l(q),k) \leq \bar{Q}_i\} \quad (40)$$

The first condition requires that expert $e_l^k$ is deployed on server $S_m$, the second condition requires that the selected expert is the target expert or a feasible substitute expert defined in (8), and the third condition requires that selecting this candidate does not exceed the per-token budget. Since each expert is deployed at least once according to (2), the target expert $e_l^{r_l(q)}$ is always deployed on some server, and its substitution quality loss is zero by (18). Therefore, the exact-execution candidate always satisfies the budget condition, which guarantees that $\mathcal{A}_{i,q,l}$ is never empty. In other words, the exact execution is not a separate fallback objective but a feasible candidate that the strategy can always fall back to when substitute execution would violate the budget. This makes the fallback behavior consistent with the global utility in (23).

### C. Look-Ahead Server-Expert Selection

To avoid moving the token to a server that is unfavorable for the following layers, the strategy estimates a look-ahead cost over the next layers. Since the future routing targets are not known before the future representations are computed, the look-ahead uses the expert transition statistics obtained on the calibration set. Let $P_l(k \to k')$ denote the empirical probability that a token routed to expert $e_l^k$ in the $l$-th MoE layer is routed to expert $e_{l+1}^{k'}$ in the next layer:

$$P_l(k \to k') = \frac{n_l(k \to k')}{\sum_{k''} n_l(k \to k'')} \quad (41)$$

where $n_l(k \to k')$ is the number of such transitions observed on the calibration set. Based on the transition statistics, the look-ahead cost-to-go of being at server $S_m$ after selecting expert $e_l^k$ in the $l$-th MoE layer is defined by a depth-$H$ recursion. The base case is $\Phi_{i,q,l}^{(0)}(m,k) = 0$, and for $h \geq 1$:

$$\Phi_{i,q,l}^{(h)}(m,k) = \sum_{k'} P_l(k \to k') \cdot \min_{(m',k') \in \mathcal{A}_{i,q,l+1}} [g_{i,q,l+1}(m',k') \mid s_{i,q,l+1} = m + \Phi_{i,q,l+1}^{(h-1)}(m',k')] \quad (42)$$

In (42), the per-step cost of the next layer is evaluated by treating the currently selected server $S_m$ as the source location $s_{i,q,l+1} = m$. Therefore, the estimated future transmission is always measured from the server chosen at the current step, which keeps the multi-hop look-ahead self-consistent with the token trajectory. The transition probability $P_l(k \to k')$ provides the expected future target distribution, so that the future cost is estimated without computing the future representations in advance.

Combining the current-layer cost and the look-ahead cost, the server-expert pair of token $q$ in the $l$-th MoE layer is selected as:

$$(m^*, k^*) = \arg \min_{(m,k)\in\mathcal{A}_{i,q,l}} [g_{i,q,l}(m,k) + \Phi_{i,q,l}^{(H-1)}(m,k)] \quad (43)$$

After the selection, the corresponding decision variable is set as $u_{i,q,l,m^*,k^*} = 1$, the token is executed on server $S_{m^*}$ by expert $e_l^{k^*}$, and the token location and the accumulated quality loss are updated as:

$$s_{i,q,l+1} = m^*, \hat{Q}_{i,q,l+1} = \hat{Q}_{i,q,l} + Q_l(r_l(q), k^*) \quad (44)$$

When the look-ahead horizon is set to $H = 1$, the look-ahead term vanishes and the selection degenerates into a myopic strategy that only minimizes the current-layer cost. A larger $H$ considers more future layers and produces a more trajectory-aware decision, at the cost of higher computation. Therefore, $H$ provides a tunable tradeoff between decision quality and runtime overhead.

*D. Weight Adaptation*

The weights $\omega_T$ and $\omega_Q$ control the tradeoff between inference delay and substitution quality loss. Since the runtime workload and network condition change over time, the weights are periodically adapted according to the observed average normalized delay $\bar{T}$ and the observed average budget utilization $\bar{Q}$ within the latest observation window:

$$\omega_Q \leftarrow \Pi_{[0,1]}(\omega_Q + \kappa\,(\bar{Q} - \bar{T})), \omega_T \leftarrow 1 - \omega_Q \quad (45)$$

where $\kappa$ is a small step size and $\Pi_{[0,1]}(\cdot)$ projects the value onto $[0,1]$. The update increases $\omega_Q$ when the budget utilization is higher than the normalized delay, so that the strategy pays more attention to quality preservation; otherwise, it increases $\omega_T$ to reduce delay. Since both $\bar{T}$ and $\bar{Q}$ are dimensionless quantities normalized in the same way as (23), the update direction is consistent with the physical meaning of the two objectives, and $\omega_T + \omega_Q = 1$ is always maintained.

*E. Online Selection Algorithm*

Based on the above models, the token-trajectory-aware online server-expert selection procedure is summarized in Table II.

Table II. Algorithm 2

Algorithm2: Quality aware token trajectory and server-expert selection

**Input:** Grouping result $\mathcal{G}$, deployment strategy $\boldsymbol{X}$, transition statistics $\{P_l\}$, look-ahead horizon $\boldsymbol{H}$, budgets $\{Q_i^{\max}\}$, network and computation parameters.

**Output:** Server-expert selection strategy $\boldsymbol{U}$ and token trajectories.

1. for each user $d_i$ and each token $q \in Q_i$ do
2. Set initial location $s_{i,q,1} = a(i)$ by (36)
3. Set accumulated quality loss $\hat{Q}_{i,q,1} = 0$ and per-token budget $\bar{Q}_i$ by (39)
4. end for
5. for each MoE layer $l = 1$ to $N$ do
6. for each user $d_i$ and each token $q \in Q_i$ do
7. Obtain target expert $r_l(q)$ from the local router
8. Build feasible candidate set $\mathcal{A}_{i,q,l}$ by (40)
9. for each candidate $(m,k) \in \mathcal{A}_{i,q,l}$ do
10. Compute per-step cost $g_{i,q,l}(m,k)$ by (37) to (38)
11. Compute look-ahead cost-to-go $\Phi_{i,q,l}^{(H-1)}(m,k)$ by the depth-$H$ recursion (42)
12. end for
13. Select $(m^*, k^*)$ with the minimum total cost by (43)
14. if $\mathcal{A}_{i,q,l}$ contains no feasible substitute under the budget then
15. Select the exact-execution candidate with $k^* = r_l(q)$
16. end if
17. Set $u_{i,q,l,m^*,k^*} = 1$
18. Execute token $q$ on server $S_{m^*}$ by expert $e_l^{k^*}$
19. Update token location $s_{i,q,l+1}$ and accumulated loss $\hat{Q}_{i,q,l+1}$ by (44)
20. end for
21. end for
22. Periodically update weights $\omega_T$ and $\omega_Q$ by (45) according to runtime observations
23. return $U$ and token trajectories.

The algorithm contains three parts.

**Line1-Line4:** Initializing token locations and quality budgets. For each token of each user, the initial location is set as the access edge server of the corresponding user, the accumulated substitution quality loss is initialized as zero, and the per-token quality budget is obtained by evenly allocating the user-level budget.

**Line5-Line21:** Performing layer-by-layer trajectory-aware selection. For each MoE layer, the target expert of each token is first obtained from the local router, and the budget-aware feasible candidate set is constructed. For each candidate, the per-step cost and the look-ahead cost-to-go are computed, where the look-ahead estimates the cost of the following layers from the expert transition statistics with the currently selected server as the source location. The candidate with the minimum total cost is then selected. If no feasible substitute candidate satisfies the budget, the exact-execution candidate is selected instead, so that the per-user quality constraint is never violated. Finally, the token is executed on the selected server by the selected expert, and the token location and the accumulated quality loss are updated.

**Line22-Line23:** Adapting weights and returning the result. The weights of inference delay and substitution quality loss are periodically updated according to the observed average normalized delay and budget utilization. Finally, the server-expert selection strategy and the token trajectories are returned.

*F. Properties of the Proposed Algorithms*

In this subsection, we analyze the properties of Algorithm 2, including feasibility, quality guarantee, trajectory consistency, and computational complexity.

**Property 1 (Feasibility).** Based on the deployment strategy generated by Algorithm 1, Algorithm 2 always selects exactly one feasible server-expert pair for each token in each MoE layer, so that constraints (c.4) to (c.7) are satisfied.

*Proof.* Consider an arbitrary token $q$ of user $d_i$ in the $l$-th MoE layer, whose target expert is $e_l^{r_l(q)}$. We first prove that the feasible candidate set $\mathcal{A}_{i,q,l}$ defined in (40) is non-empty, and then prove that the selection satisfies (c.4) to (c.7).

1) *Non-emptiness.* According to constraint (2), every expert is deployed on at least one edge server. Hence, there exists at least one server $S_{m_0}$ with $x_{m_0,l,r_l(q)} = 1$. For the candidate $(m_0, r_l(q))$, the three conditions of $\mathcal{A}_{i,q,l}$ in (40) are checked as follows. The first condition $x_{m_0,l,r_l(q)} = 1$ holds by the above deployment. The second condition holds because the target expert itself belongs to its own feasible execution expert set, i.e., $e_l^{r_l(q)} \in V_{l,r_l(q)}$ by the definition in (8). The third condition holds because, by the substitution quality model in (18), the substitution quality loss of the target expert is $Q_l(r_l(q), r_l(q)) = 0$, so that $\hat{Q}_{i,q,l} + Q_l(r_l(q), r_l(q)) = \hat{Q}_{i,q,l} \le \bar{Q}_i$, where the last inequality follows from the budget invariant maintained by (44), which is proved in Property 2. Therefore, the exact-execution candidate $(m_0, r_l(q))$ always belongs to $\mathcal{A}_{i,q,l}$, and $\mathcal{A}_{i,q,l} \ne \emptyset$.

2) *Feasibility of the selection.* Since $\mathcal{A}_{i,q,l}$ is non-empty, the minimization in (43) returns a well-defined pair $(m^*, k^*) \in \mathcal{A}_{i,q,l}$. By the membership in $\mathcal{A}_{i,q,l}$, we have $x_{m^*,l,k^*} = 1$, which satisfies the deployment constraint (c.6), and $e_l^{k^*} \in V_{l,r_l(q)}$, which satisfies the feasible-expert constraint (c.7). The algorithm sets exactly one decision variable $u_{i,q,l,m^*,k^*} = 1$ and leaves all the other variables of this token and layer as zero, which gives $\sum_m \sum_k u_{i,q,l,m,k} = 1$ and satisfies the binary constraint (c.4) and the single-selection constraint (c.5).

Since the above holds for an arbitrary token in an arbitrary MoE layer, Algorithm 2 always produces a feasible solution. ■

**Property 2 (Quality Guarantee).** Algorithm 2 always satisfies the per-user quality constraint (c.8), i.e., $Q_i \le Q_i^{\max}$ for every user $d_i \in \mathcal{D}$.

*Proof.* We first prove a per-token budget invariant by induction over the MoE layers, and then sum over the tokens of each user.

1) *Budget invariant.* For token $q$ of user $d_i$, we claim that the accumulated substitution quality loss satisfies $\hat{Q}_{i,q,l} \le \bar{Q}_i$ before every MoE layer $l$, and that the final accumulated loss satisfies $\hat{Q}_{i,q,N+1} \le \bar{Q}_i$.

*Base case.* Before the first MoE layer, the accumulated loss is initialized as $\hat{Q}_{i,q,1} = 0 \le \bar{Q}_i$, since $\bar{Q}_i = Q_i^{\max}/|Q_i| \ge 0$ by (39).

*Inductive step.* Assume $\hat{Q}_{i,q,l} \le \bar{Q}_i$ holds before the $l$-th MoE layer. The selected pair $(m^*, k^*)$ belongs to $\mathcal{A}_{i,q,l}$, so the third condition of (40) gives $\hat{Q}_{i,q,l} + Q_l(r_l(q), k^*) \le \bar{Q}_i$.

By the update rule in (44), the accumulated loss before the next layer is $\hat{Q}_{i,q,l+1} = \hat{Q}_{i,q,l} + Q_l(r_l(q), k^*) \le \bar{Q}_i$. Therefore, the invariant holds before the $(l+1)$-th MoE layer. By induction, it holds for all $l = 1, \ldots, N$, and in particular $\hat{Q}_{i,q,N+1} \le \bar{Q}_i$.

2) *Per-token loss bound.* The final accumulated loss is the sum of the per-layer substitution quality losses of the token, that is $\sum_{l=1}^{N} Q_{i,q,l} = \hat{Q}_{i,q,N+1} \le \bar{Q}_i$, where $Q_{i,q,l} = Q_l(r_l(q), k^*)$ is the realized per-layer loss in (18) under the selected pair.

3) *User-level bound.* Summing the per-token bound over all tokens of user $d_i$ and using the per-token budget allocation in (39), the user-level substitution quality loss in (20) satisfies $Q_i = \sum_{q \in Q_i} \sum_{l=1}^{N} Q_{i,q,l} \le \sum_{q \in Q_i} \bar{Q}_i = |Q_i| \cdot \frac{Q_i^{\max}}{|Q_i|} = Q_i^{\max}$.

Therefore, the per-user quality constraint (c.8) in (23) is satisfied for every user. ■

**Property 3 (Trajectory Consistency and Look-Ahead Optimality).** The look-ahead estimation in (42) is consistent with the realized token trajectory, and when the calibration transition statistics match the runtime routing distribution, the look-ahead selection in (43) is no worse than the myopic selection ($H = 1$) in expectation.

*Proof.* We prove the two parts separately.

1) *Trajectory consistency.* In the depth-$H$ recursion (42), the per-step cost of the next layer is evaluated under the condition $s_{i,q,l+1} = m$, where $S_m$ is the server selected at the current step. According to the token-trajectory model in (36), the realized source location of the $(l+1)$-th layer equals the execution server of the $l$-th layer, i.e., $s_{i,q,l+1} = m^*$ once $u_{i,q,l,m^*,k^*} = 1$ is set by (44). Therefore, for the pair $(m^*, k^*)$ that is finally selected, the source location assumed in the look-ahead is exactly the realized source location of the token. As a consequence, the transmission term $\mathbb{1}(s_{i,q,l+1} \ne m')(z_{l+1}/B_{s_{i,q,l+1},m'} + \tau_{s_{i,q,l+1},m'})$ in (37) is computed from a reachable location, so the look-ahead never assumes a token position that the trajectory cannot reach and never underestimates the future transmission by an unreachable shortcut.

2) *Expected optimality with respect to the myopic strategy.* Let $J_{i,q,l}^{(H)}(m,k) = g_{i,q,l}(m,k) + \Phi_{i,q,l}^{(H-1)}(m,k)$, which denote the total cost used in (43), and let $J_{i,q,l}^{(1)}(m,k) = g_{i,q,l}(m,k)$ denote the myopic cost, obtained by setting $H = 1$ so that the look-ahead term $\Phi_{i,q,l}^{(0)}(m,k) = 0$ vanishes. Define the expected realized cost-to-go of selecting $(m, k)$ as the expectation of the per-step costs of the following layers along the trajectory induced by (36), where the expectation is taken over the routing targets of the future layers.

By definition (41), the calibration transition probability $P_l(k \to k')$ is the empirical frequency of the routing transition $e_l^k \to e_{l+1}^{k'}$. When this empirical distribution matches the runtime routing distribution, the recursion (42) computes, for each future layer, the expectation of the minimum feasible per-step cost over the true target distribution, with the source location fixed to the server selected at the previous step as shown in part (i). Hence $\Phi_{i,q,l}^{(H-1)}(m,k)$ is an unbiased estimate of the expected realized cost of the next $H-1$ layers conditioned on selecting $(m, k)$ and on acting greedily within the budget thereafter.

Let $\left(m^*, k^*\right) = \arg\min_{(m,k) \in \mathcal{A}_{i,q,l}} J_{i,q,l}^{(H)}(m,k)$ be the look-ahead choice and $\left(\tilde{m}, \tilde{k}\right) = \arg\min_{(m,k) \in \mathcal{A}_{i,q,l}} J_{i,q,l}^{(1)}(m,k)$ be the myopic choice. Since $(m^*, k^*)$ minimizes the total cost over the same feasible set $\mathcal{A}_{i,q,l}$ that contains $\left(\tilde{m}, \tilde{k}\right)$, we have $J_{i,q,l}^{(H)}(m^*, k^*) \le J_{i,q,l}^{(H)}(\tilde{m}, \tilde{k})$.

Taking expectations over the future routing targets and using the unbiasedness of $\Phi^{(H-1)}$, the expected current-plus-future cost of the look-ahead choice is no larger than that of the myopic choice, because the myopic choice ignores the non-negative future term and may therefore commit to a server that induces a higher expected future transmission. Therefore, the look-ahead selection is no worse than the myopic selection in expectation. When all candidates share the same expected future cost, the two selections coincide, which shows that the improvement is non-strict in general. ■

**Property 4 (Polynomial-Time Complexity).** Algorithm 2 has polynomial-time complexity in the number of users, tokens, MoE layers, edge servers, and experts for any fixed look-ahead horizon $H$.

*Proof.* We count the operations of Algorithm 2 from the innermost computation outward.

1) *Candidate-set construction.* For a single token in the $l$-th MoE layer, building $\mathcal{A}_{i,q,l}$ in (40) checks, for each server $S_m$ and each feasible expert $e_l^k \in V_{l,r_l(q)}$, the deployment, feasibility, and budget conditions in constant time. Since the feasible execution expert set $V_{l,r_l(q)}$ is restricted to one expert group by (8) and contains at most $\bar{E}$ experts on average, the candidate set is built in $O(M\bar{E})$ operations.

2) *Per-step cost.* Each per-step cost $g_{i,q,l}(m,k)$ in (37) and (38) is computed in constant time, so evaluating all candidates of one step takes $O(M\bar{E})$ operations.

3) *Look-ahead recursion.* The depth- $H$ recursion (42) expands each candidate into at most $M\bar{E}$ next-layer candidates, weighted by the transition probabilities in (41). Therefore, the number of expanded paths is bounded by $O((M\bar{E})^{H-1})$, and evaluating the recursion for one candidate of the current step costs $O((M\bar{E})^{H-1})$. Combining with step (i), the total cost of selecting the pair for one token in one MoE layer is $O(M\bar{E} \cdot (M\bar{E})^{H-1}) = O((M\bar{E})^H)$.

4) *Total complexity.* The selection is repeated for each MoE layer of each token of each user. Since there are $N$ MoE layers and $\sum_{i=1}^{U} |Q_i|$ tokens in total, the overall complexity of Algorithm 2 is $O(N\sum_{i=1}^{U} |Q_i| \cdot (M\bar{E})^H)$.

The periodic weight update in (45) is performed once per observation window and costs $O(1)$ per update, so it does not change the dominant term. Because the horizon $H$ is a small constant in practice and the candidate set is restricted to one expert group, the factor $(M\bar{E})^H$ is bounded by a low-degree polynomial. Therefore, Algorithm 2 runs in polynomial time and can be executed in real time. ■

## VI. Performance Evaluation

In this section, we evaluate and analyze the performance of the proposed OrderMoE framework in detail.

### *A. Experimental Setup*

**Experimental environment:** In this paper, we implement the proposed OrderMoE framework based on PyTorch and deploy it on a real distributed edge testbed consisting of 8 physical edge servers. The servers are equipped with different GPUs, and their computation capacities $F_m$ range from 20 to 110 TFLOPS, while their memory capacities $C_m$ range from 12 to 48 GB. The servers are connected by heterogeneous links whose bandwidths $B_{m,n}$ range from 1 to 10 Gbps and whose propagation latencies $\tau_{m,n}$ range from 0.2 to 20 $ms$, so that both computation and communication heterogeneity are captured. Unless otherwise specified, the user requests arrive at a rate of 40 requests per second and are associated with the edge servers according to a non-uniform spatial distribution.

**Models and tasks.** We evaluate OrderMoE on three representative MoE models, namely Switch-Base-8E, Qwen-MoE-A2.7B, and Mixtral-8x7B, which cover small, medium, and large scales. The inference quality is measured on WikiText-103 (perplexity), SQuAD (F1 score), and GSM8K (accuracy). The expert similarity, the activation frequency, and the expert transition statistics are obtained on a held-out calibration set. Unless otherwise specified, the look-ahead horizon is set to $H = 3$, the layer-aware similarity threshold is set with $\theta_{\min} = 0.5$ and $\theta_{\max} = 0.9$, and the memory budget ratio is set to 2.0.

**Baselines.** We compare OrderMoE with eleven methods, including four representative distributed inference frameworks (Edge-LLM, EdgeShard, WDMoE, and MoE²), four deployment or routing baselines (Random Deployment, Popularity-based Replication, Similarity Deployment without Substitute, and Local Substitute Only), and two internal variants of OrderMoE (OrderMoE without Look-Ahead and OrderMoE without Redundancy). The performance metrics include the average inference latency, the cross-server communication volume, the inference throughput, the expert execution type breakdown, and the inference quality.

### *B. Overall Performance Comparison*

In this subsection, we compare the overall performance of OrderMoE with different baselines on Mixtral-8x7B. The results are shown in Fig.4, where Fig.4(a) presents the average inference latency, Fig.4(b) presents the cross-server communication volume per one thousand tokens, and Fig.4(c) presents the inference throughput.

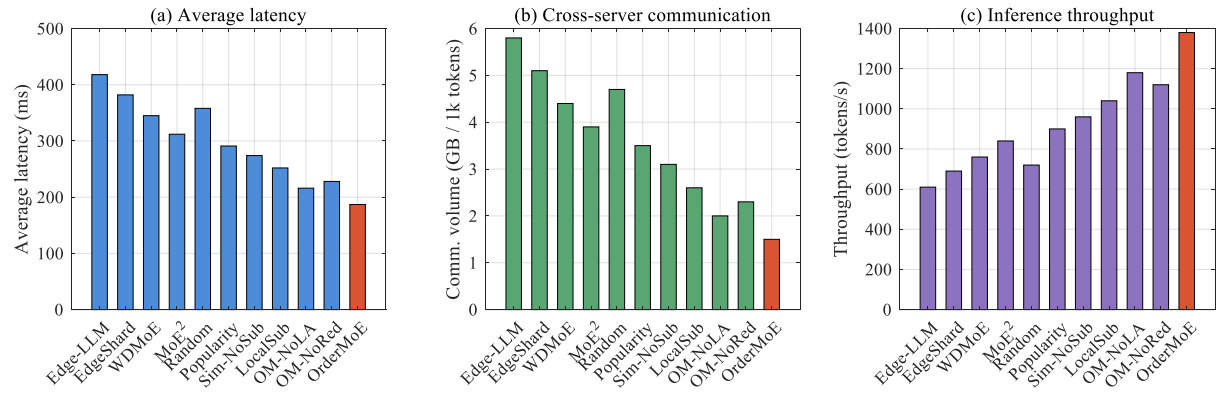

Fig.4. Overall performance comparison

rom Fig.4(a), the full OrderMoE achieves the lowest average latency of 187ms, which is 51.0% lower than EdgeShard and 40.1% lower than MoE². This is because OrderMoE disperses similar experts across servers and selects nearby exact or substitute experts, so that most tokens are executed locally and the cross-server transmission is greatly reduced. From Fig.4(b), the cross-server communication of OrderMoE is only 1.5 GB per one thousand tokens, which is the lowest among all methods, and from Fig.4(c), OrderMoE achieves the highest throughput of 1380 tokens per second. The two internal variants, OrderMoE without Look-Ahead and OrderMoE without Redundancy, perform worse than the full OrderMoE but still better than the other baselines, which confirms that both the look-ahead selection and the redundant deployment contribute to the overall performance.

### *C. Execution Behavior and Inference Quality*

In this subsection, we evaluate the execution behavior and the inference quality of different methods. The results are shown in Fig.5. Fig.5(a) presents the expert execution type breakdown, including local exact execution, local substitute execution, remote exact execution, and remote substitute execution. Fig.5(b) presents the inference quality on WikiText-103, SQuAD, and GSM8K.

From Fig.5(a), the full OrderMoE achieves the highest local execution ratio of 83%, where the local exact execution accounts for 52% and the local substitute execution accounts for 31%. In contrast, Similarity Deployment without Substitute cannot perform any substitute execution, so its remote exact execution reaches 62%, which directly increases the cross-server transmission. From Fig.5(b), the inference quality of OrderMoE is very close to the methods that always perform exact execution. Specifically, the perplexity on WikiText-103 increases by only 0.3, the F1 score on SQuAD decreases by only 0.6, and the accuracy on GSM8K decreases by only 0.7. This is because the substitute execution is restricted by the layer-aware similarity constraint and the per-user quality budget, so the substitution quality loss is well bounded.

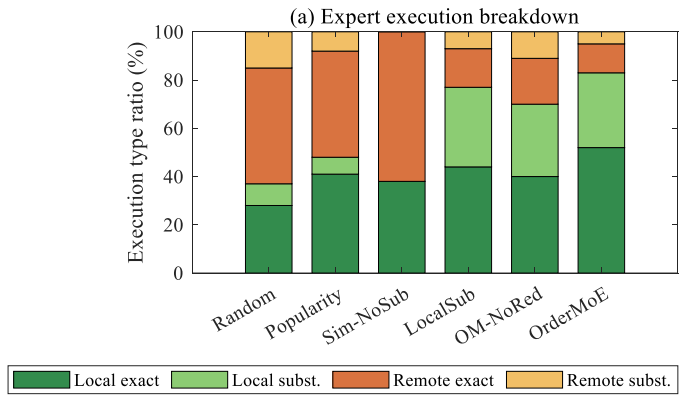

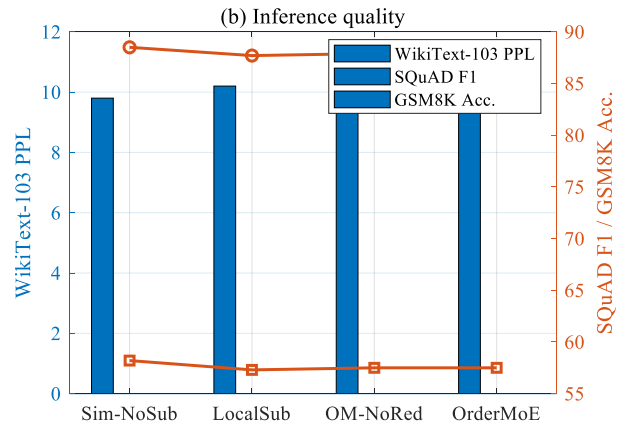

Fig.5. Execution behavior and inference quality

*D. Comparison with Related Work Baselines on Different Models*

In this subsection, we evaluate the performance of OrderMoE and the related work baselines on different MoE models, including Switch-Base-8E, Qwen-MoE-A2.7B, and Mixtral-8x7B. The results are shown in Fig.6, where Fig.6(a) presents the average latency and Fig.6(b) presents the quality retention ratio with respect to the full centralized model.

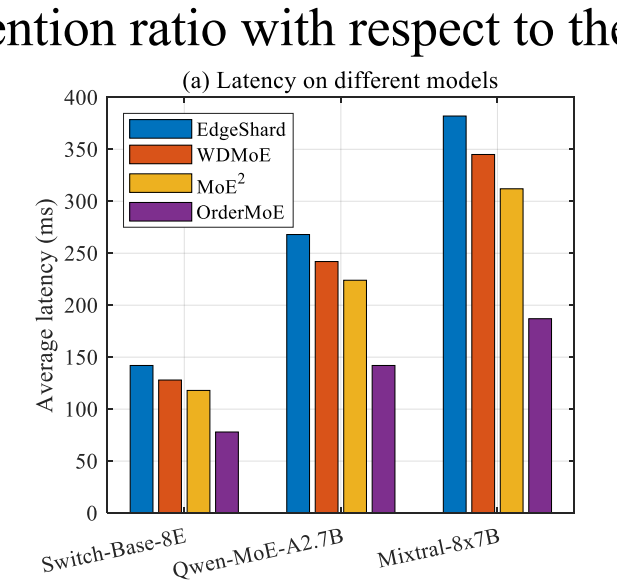

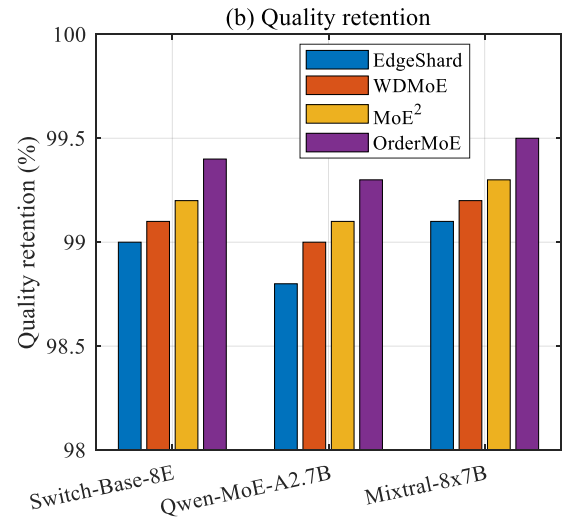

Fig.6. Comparison on different MoE models

From Fig.6(a), OrderMoE consistently achieves the lowest latency on all three models, and the latency reduction becomes larger as the model scale increases. On the largest model Mixtral-8x7B, OrderMoE reduces the latency by 40.1% compared with MoE². This is because larger models have more experts, which provides more grouping and substitution opportunities for OrderMoE. From Fig.6(b), the quality retention of OrderMoE is above 99.3% on all models, which confirms that OrderMoE preserves the inference quality while reducing the latency.

*E. Effect of the Look-Ahead Horizon*

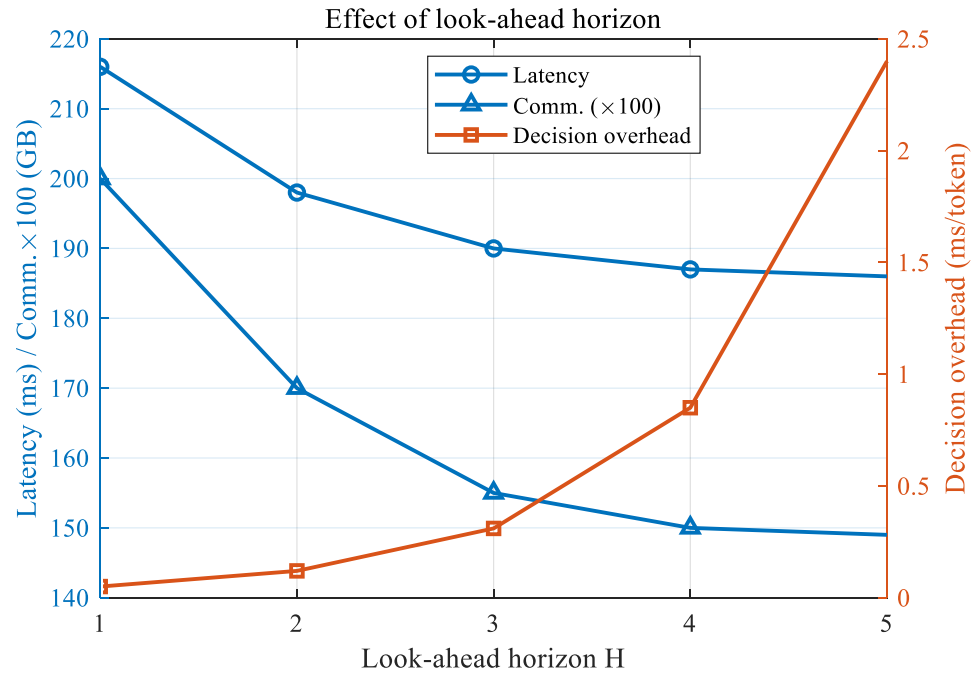

Fig.7. Effect of inter-server bandwidth

In this subsection, we evaluate the effect of the look-ahead horizon $H$ on Mixtral-8x7B. The horizon is varied from 1 to 5, where $H = 1$ corresponds to the myopic selection. The results are shown in Fig.7.

From Fig.7, the average latency and the cross-server communication decrease quickly when $H$ increases from 1 to 3, because considering more future layers helps the strategy avoid moving the token to a server that is unfavorable for the following layers. When $H$ exceeds 3, the latency almost saturates, while the per-token decision overhead grows rapidly because the look-ahead recursion expands more candidate paths. Therefore, $H = 3$ provides a good tradeoff between decision quality and runtime overhead, which is consistent with the complexity analysis in Property 4.

*F. Effect of the Similarity Threshold*

In this subsection, we evaluate the effect of the similarity threshold on Mixtral-8x7B. For the fixed-threshold scheme, the threshold $\theta$ is varied from 0.5 to 0.9, and the resulting latency-quality tradeoff is compared with the proposed layer-aware threshold. The results are shown in Fig.8.

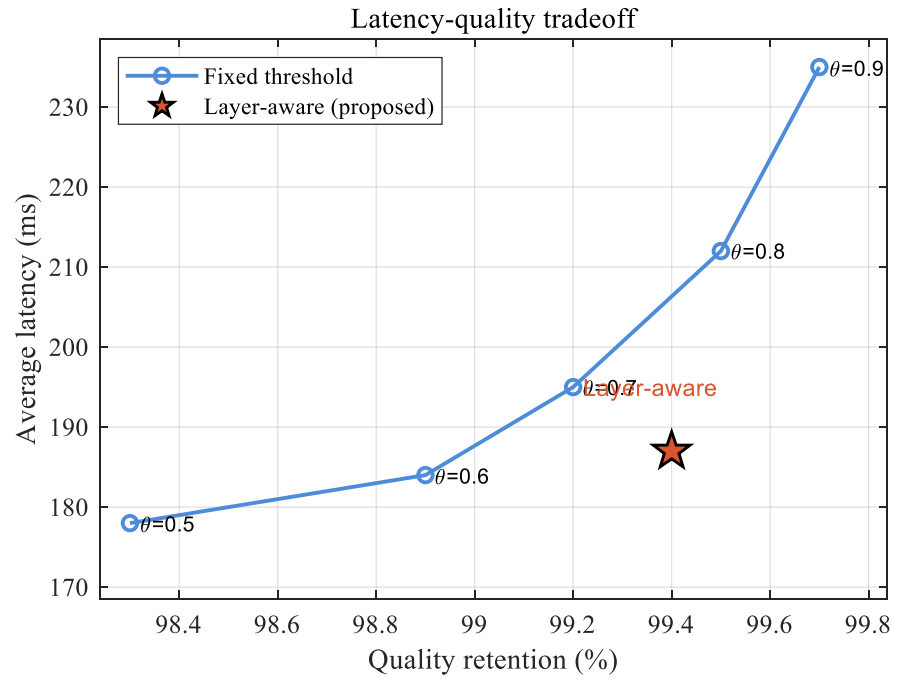

Fig.8. Effect of expert memory budget

From Fig.8, a smaller fixed threshold allows more substitute experts and therefore reduces the latency, but it also increases the substitution quality loss. A larger fixed threshold preserves the quality but reduces the substitution opportunities and increases the latency. The proposed layer-aware threshold lies below the fixed-threshold tradeoff curve, which means that it achieves a lower latency at the same quality level. This is because the layer-aware threshold allows more substitution in deep layers where experts are more similar, and restricts substitution in shallow layers where experts are more specialized.

*G. Effect of the Redundant Deployment*

In this subsection, we evaluate the effect of the redundant deployment on Mixtral-8x7B. The memory budget ratio, defined as the total deployed memory capacity over the memory of a single full model replica, is varied from 1.0 to 3.0. The proposed importance-aware redundant deployment is compared with the popularity-based replication. The results are shown in Fig.9, where Fig.9(a) presents the local group coverage ratio and Fig.9(b) presents the average latency.

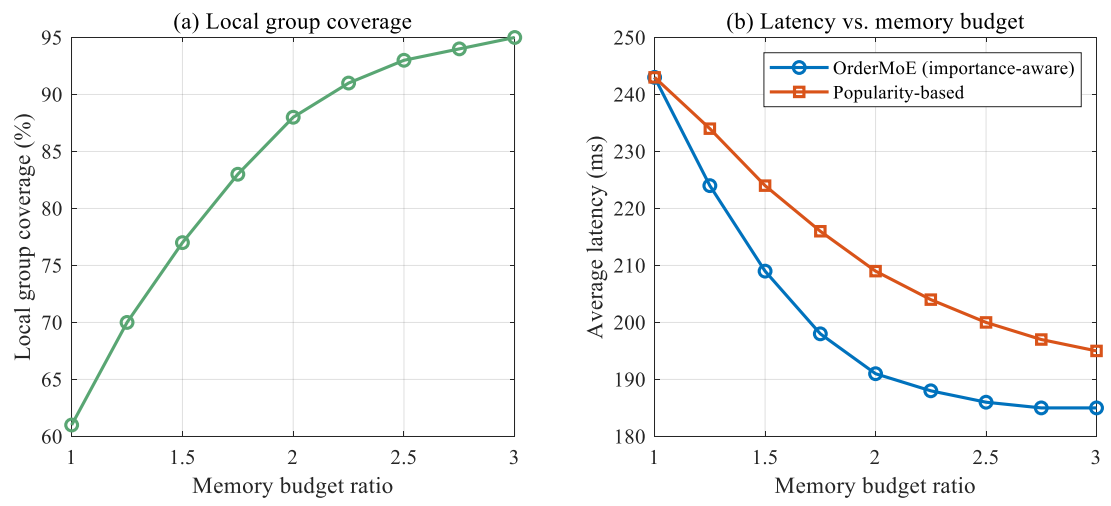

Fig.9. Latency quality tradeoff

From Fig.9(a), the local group coverage of OrderMoE

increases quickly as the memory budget increases, and reaches 88% at a memory budget ratio of 2.0. From Fig.9(b), the latency of OrderMoE decreases faster than the popularity-based replication under the same memory budget. This is because the proposed redundancy benefit prefers the replica that covers an uncovered group of an important expert on a busy server with the highest benefit per unit memory, so the limited memory is used more efficiently to extend the local coverage.

*H. Ablation Study*

In this subsection, we conduct an ablation study on Mixtral-8x7B to evaluate the contribution of each component, including the look-ahead selection, the redundant deployment, the substitute execution, and the same-group dispersion. The results are shown in Fig.10, where Fig.10(a), Fig.10(b), and Fig.10(c) present the average latency, the cross-server communication, and the quality degradation, respectively.

From Fig.10(a) and Fig.10(b), removing any component increases both the latency and the communication. Removing the substitute execution causes the largest latency increase, because the tokens can only be executed by the exact experts, which often reside on remote servers. Removing the same-group dispersion also increases the latency, because the similar experts may be placed on the same server and the local substitution opportunities are reduced. Fig.10(c) shows that the quality degradation of all OrderMoE variants is small. This is because all variants still follow the expert similarity constraint and the per-user quality budget. Therefore, the main function of the removed components is to improve latency and communication efficiency rather than to directly preserve the quality.

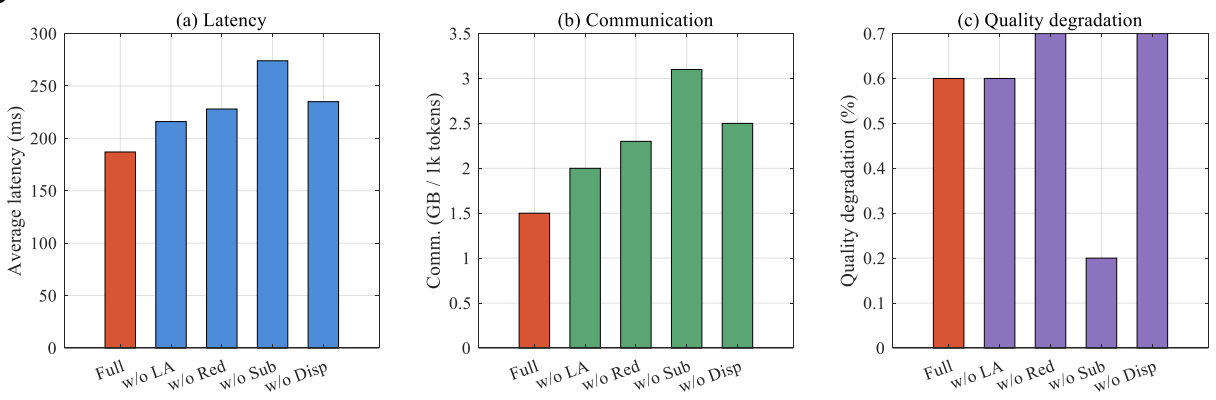

Fig.10. Ablation study

*I. Performance Under Different Request Loads*

In this subsection, we evaluate the performance of different methods under different request loads. The request load is varied from 10 to 80 requests per second. The results are shown in Fig.11, where Fig.11(a) presents the average latency and Fig.11(b) presents the SLA satisfaction ratio, defined as the ratio of requests whose latency is below the target.

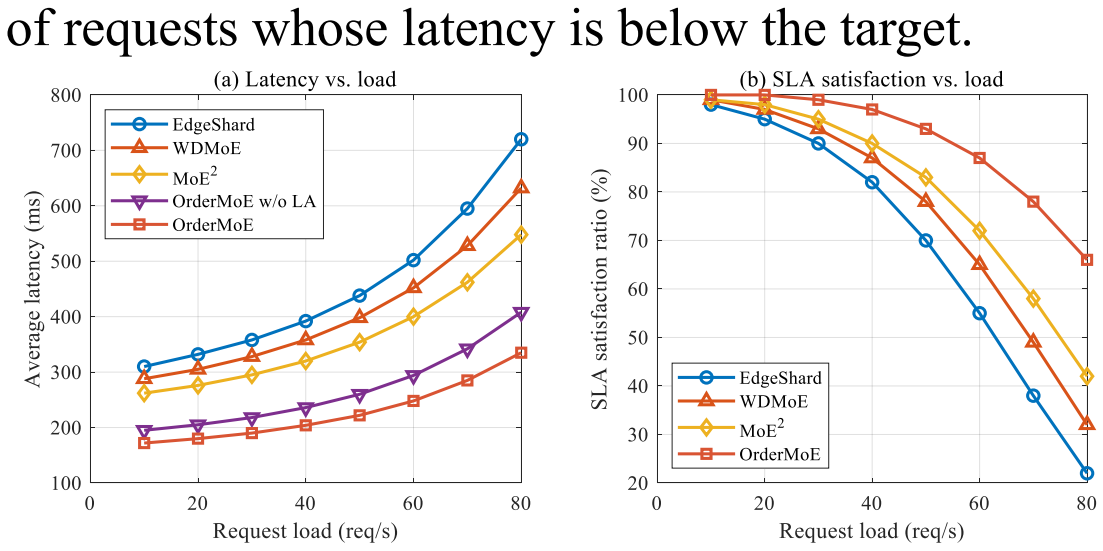

Fig.11. Effect of request load

From Fig.11(a), the average latency increases when the request load increases. This is because a higher request load increases the GPU queueing delay and the link contention. However, the latency increase of OrderMoE is smaller than those of the baselines. This is because OrderMoE reduces the cross-server communication and distributes the expert execution more flexibly across the edge servers. From Fig.11(b), the SLA satisfaction ratio of OrderMoE remains above 66% even at 80 requests per second, while the baselines drop quickly under heavy load. This confirms that OrderMoE is more robust to the request load.

*J. Effect of Number of Edge Servers and Look-Ahead Window*

In this subsection, we evaluate the scalability and the runtime decision overhead of OrderMoE on Mixtral-8x7B. The results are shown in Fig.12, where Fig.12(a) presents the average latency under different numbers of edge servers, and Fig.12(b) presents the per-token decision overhead and the effective end-to-end latency under different look-ahead window sizes. To assess scalability beyond the physical testbed, the number of edge servers is scaled from 4 to 16 by emulating additional servers with the same heterogeneity profile, while the aggregate request load is kept proportional to the number of servers.

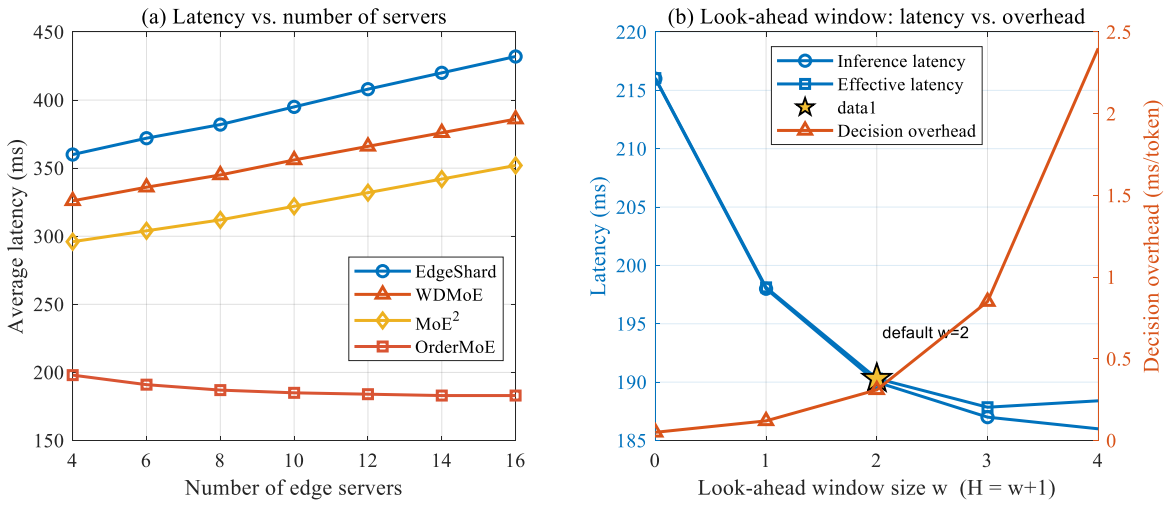

Fig.12. Server scalability and look ahead window

From Fig.12(a), the latency of the exact-execution baselines increases steadily as the number of servers grows, rising by about 20% for EdgeShard and about 19% for MoE² when the server count goes from 4 to 16. The cause lies in the source-dependent transmission term $z_l/B_{s,m} + \tau_{s,m}$ in (37): for an exact-execution method, a token must reach the unique server that holds its target expert, and as the experts are spread over more servers, the probability that the target resides on a remote server grows roughly with the server count. More servers therefore mean more remote exact invocations and a larger accumulated transmission delay, so scaling out paradoxically hurts these baselines. In contrast, the latency of OrderMoE stays essentially flat, even decreasing slightly from 198 ms at 4 servers to a stable 183 ms beyond 12 servers. Two mechanisms explain this stability. First, the similarity-aware grouping with same-group dispersion in (30) guarantees that every server hosts a feasible in-group expert for most groups, so a token can be served locally by substitution regardless of how many servers exist, and the remote probability does not grow with the server count. Second, adding servers brings additional aggregate memory, which the importance-aware redundancy in (33) immediately spends on covering more groups locally, so the marginal servers improve coverage rather than dispersing targets out of reach. The slight early decrease reflects this coverage gain at small scale, and the flattening reflects that once coverage is near-complete, extra memory yields diminishing benefit. This decoupling of latency from the server count is the property that makes OrderMoE scalable, whereas the baselines degrade precisely because they couple execution to a fixed expert location.

From Fig.12(b), increasing the look-ahead window from $w = 0$ to $w = 2$ reduces the average inference latency from 216 ms to 190 ms, because a wider window lets the self-consistent recursion in (42) anticipate the multi-hop transmissions induced by the trajectory coupling in (36) and steer the token onto servers that remain cheap for the following layers. Beyond $w = 2$, however, the inference latency barely improves while the per-token decision overhead climbs from 0.31 ms to 2.40 ms, because the recursion expands $O((M\bar{E})^w)$ candidate paths as established in Property 4, and the additional layers carry increasingly uncertain, discounted transition information. The effective end-to-end latency, which adds the decision overhead to the inference latency, therefore forms a U-shape: it falls from 216.1 ms at $w = 0$ to a minimum of 190.3 ms at $w = 2$, then rises to 188.4 ms at $w = 4$ as the overhead outweighs the shrinking inference gain. This U-shape is the operational reason for choosing $w = 2$ (equivalently $H = 3$) as the default: it sits exactly at the knee where the marginal transmission saving is overtaken by the marginal computation cost of the look-ahead itself, which is fully consistent with both the horizon study in Fig.7 and the complexity bound in Property 4.

## VII. Conclusion

In this paper, we recognize a gap in prior distributed edge MoE inference research, which has predominantly focused on optimizing exact expert access through caching, replication, prefetching, and communication scheduling, while neglecting the impact of expert similarity, a critical factor for reducing cross-server token transmission under bandwidth-limited edge environments. To accelerate MoE inference at the edge and to strike a balance among inference latency, communication overhead, server workload, and inference quality, we introduce a similarity-aware expert grouping and distributed deployment framework for edge MoE inference, or OrderMoE for short. OrderMoE distinctively considers expert similarity, substitute execution, expert deployment, and token trajectory to identify efficient expert placement and runtime server-expert selection strategies. Specifically, OrderMoE constructs a router-induced logits based expert similarity model and groups experts in each MoE layer according to their functional similarity. Based on the grouping results, it develops a similarity-aware distributed deployment strategy, where representative experts from different similarity groups are placed across edge servers and important experts are redundantly deployed according to their activation frequency and similarity role. Given that reducing remote expert invocation and maintaining inference quality are conflicting objectives, we further design a quality-aware and trajectory-aware runtime selection algorithm to determine whether a token should be processed by a local substitute expert or forwarded to a remote target expert. The algorithm also updates token locations across MoE layers to reduce future movement cost. The experimental results on a real distributed edge testbed indicate that OrderMoE significantly outperforms existing methods in terms of average latency, tail latency, cross-server communication traffic, and remote exact expert invocation ratio, while maintaining small and controllable inference quality degradation.

## Acknowledgment

This work was supported in part by the grant from NSFC Grant no. 62571156, 62101159, 52475009, NSF of Shandong Grant no. ZR2021MF055, ZR2025QC666, the Research Grants Council of Hong Kong under the Areas of Excellence scheme grant AoE/E-601/22-R, and also the Opening Project of the Key Laboratory of Advanced Manufacturing and Intelligent Technology (Ministry of Education) at Harbin University of Science and Technology (KFKT202306).

Additionally, the authors used an AI-based language assistance tool to improve the clarity and readability of parts of the manuscript, particularly the Abstract and Introduction, and all technical content, analysis, and conclusions were developed and carefully verified by the authors.